\documentclass[arxiv,nonblindrev]{informs3} 
\usepackage{tikz}
\usepackage{amsmath}
\usepackage{amsfonts}
\usetikzlibrary{shapes.geometric, arrows, shadows}
\usetikzlibrary{fit,backgrounds}
\usetikzlibrary{shadows.blur}
\usetikzlibrary{arrows.meta, positioning, shapes.geometric}

\usepackage{tikzscale}
\usepackage{pdfpages}
\usepackage{verbatim}  
\usepackage{vector}  
\usepackage{amssymb}
\usepackage{bbm}
\usepackage{tabularx} 
\usepackage{amsmath}
\usepackage{braket}
\usepackage{mathtools}
\usepackage{xcolor}
\usepackage{float}
\usepackage{graphicx}
\usepackage{caption}
\usepackage{subcaption}
\usepackage{pdflscape}
\usepackage[section]{placeins}

\usepackage[linesnumbered,ruled,vlined]{algorithm2e}

\DeclarePairedDelimiter{\norm}{\lVert}{\rVert}

\OneAndAHalfSpacedXI

\usepackage[numbers]{natbib}
 \bibpunct[, ]{(}{)}{,}{a}{}{,}%
 \def\bibfont{\small}%
\usepackage{hyperref}%
\hypersetup{
  colorlinks,
  citecolor=blue,
  linkcolor=blue,
  urlcolor=blue
  }
\TheoremsNumberedThrough     

\EquationsNumberedBySection 

\begin{document}

\YEAR{2024}
\FIRSTPAGE{000}%

\RUNAUTHOR{Katial, Smith-Miles, and Hill}

\RUNTITLE{On the Instance Dependence of Optimal Parameters for the QAOA: Insights via ISA}

\TITLE{On the Instance Dependence of Optimal Parameters for the Quantum Approximate Optimisation Algorithm: Insights via Instance Space Analysis}

\ARTICLEAUTHORS{%
\AUTHOR{Vivek Katial, Kate Smith-Miles}
\AFF{School of Mathematics and Statistics, The University of Melbourne, Australia \EMAIL{\\vkatial@student.unimelb.edu, smith-miles@unimelb.edu.au} \URL{}}
\AUTHOR{Charles Hill}
\AFF{School of Physics, University of New South Wales, Australia \EMAIL{\\charles.hill1@unsw.edu.au}}
} 

\ABSTRACT{%
The performance of the Quantum Approximate Optimisation Algorithm (QAOA) relies on the setting of optimal parameters in each layer of the circuit. This is no trivial task, and much literature has focused on the challenge of finding optimal parameters when the landscape is plagued with problems such as "barren plateaus". There are many choices of optimisation heuristics that can be used to search for optimal parameters, each with its own parameters and initialisation choices that affect performance. More recently, the question of whether such optimal parameter search is even necessary has been posed, with some studies showing that optimal parameters tend to be concentrated on certain values for specific types of problem instances. However, these existing studies have only examined specific instance classes of MaxCut, so it is uncertain if the claims of instance independence apply to a diverse range of instances. In this paper, we use Instance Space Analysis to study the dependence of instance characteristics on the performance of QAOA. Focusing on the MaxCut problem, we assess the effectiveness of parameter initialisation strategies and introduce a new initialisation approach based on instance characteristics called Quantum Instance-Based Parameter Initialisation (QIBPI). This study reveals that using insights about instance characteristics in choosing initialisation parameters can improve QAOA performance. We also show that, within certain instance classes, parameters from smaller instances can be transferred to larger ones. This research provides a foundation for further instance space analysis for quantum algorithms and encourages a broader class of instances to be considered to ensure conclusions are not limited to particular well-studied test problems or classes.
}%



\maketitle

%
\section{Introduction}\label{intro} 
The field of quantum computing has attracted much attention in recent years due to its potential to evaluate a combinatorial and exponentially growing search space more efficiently than mere brute force enumeration using classical computers. Despite some success achieved in recent years \citep{Bravyi2018}, \citep{Arute2019} where quantum algorithms demonstrated some advantage in the form of quantum speed-ups for certain applications, the ability of quantum computers to solve large-scale optimisation problems is still a promise yet to be fulfilled \citep{abbas2023quantum}.  The challenges of demonstrating advantage include firstly the design of quantum algorithms that can consistently find feasible and near-optimal solutions to optimisation problems, made even more difficult for constrained optimisation problems where common quantum algorithms handle constraints as penalty terms alongside an objective function so that the problem can be mapped onto a Quadratic Unconstrained Binary Optimisation (QUBO) framework \citep{lucas2014ising,glover2022quantum}, however, this may not be a natural formulation and creates challenges for parameter optimisation to ensure the global minima of the energy landscape align with feasible and optimal solutions to the optimisation problem. Secondly, even if a quantum algorithm can find near-optimal solutions to some instances of the problem, this does not mean it will be an effective strategy for all instances of the problem, and little is currently known about the dependence of quantum algorithm performance on instance characteristics. Thirdly, the concept of quantum advantage relies on demonstrating that an effective quantum algorithm can find solutions in a manner that scales better than conventional computers, but this involves acknowledging that conventional computers are rarely used to solve optimisation problems using brute force enumeration. Instead, effective mathematical techniques have been developed over many decades to enable the search space to be pruned, such that state-of-the-art classical algorithms can often solve very large-scale optimisation problems in a manner that sets a high bar for improvement by quantum computers.
\\\\
Given these challenges, in this paper, we demonstrate how the recently developed methodology known as Instance Space Analysis (ISA) \citep{smith2023instance} can be used to provide useful insights to drive future research into quantum algorithms for optimisation problems. Specifically, we show how ISA can be used to address the abovementioned first two challenges, by exploring how parameter initialisation of the Quantum Approximate Optimisation Algorithm (QAOA) affects the performance of the algorithm, and simultaneously how this interplays with the characteristics of the test instance that create the energy landscape. We select as a case study the unconstrained MaxCut problem, as one of the more promising benchmark problems considering the third abovementioned challenge.
\\\\
Most studies of QAOA have focused on the $\text{MaxCut}$ problem, however, the family of graphs studied has often been limited to a specific few, namely $d$-regular graphs and Erdős–Rényi (also known as Uniform Random) graphs \citep{Basso2022}, \citep{Wurtz2021}, \citep{Rancic2023}, \citep{Farhi2014}. QAOA is a parameterised algorithm that is characterised by parameters $(\gamma, \beta)$. A critical task of QAOA is determining the optimal parameters of $(\gamma, \beta)$ that minimise an objective function. \cite{Brandao2018} presented findings that for 3-regular graphs, QAOA exhibits the \textit{concentration of parameters}. Essentially, for such cases, the optimal setting of parameters do not depend on the individual problem instance. As such, once optimal parameters for one problem have been identified, it is possible to apply them to other instances, even if those instances are larger in scale.  Such a finding has the ability to significantly reduce calls to the quantum device \textit{and} reduce the number of iterations in the classical optimisation routine. The result has been reproduced in many studies: \cite{Zhou2020}, \cite{Lee2021}, \cite{Streif2019}, \cite{Galda2021}. However, because the same families of instances have been studied, there remains a paucity of evidence suggesting that this phenomenon extends to all instances. \cite{Herrman2021} first studied the impact of graph structure on algorithm performance, where they identified the existence of \textit{odd-cycles} and the amount of symmetry in a $\text{MaxCut}$ instance as a strong predictor of QAOA success. However, this work did not investigate the efficacy of different parameter initialisation techniques based on graph structure. Outside of $\text{MaxCut}$, prior work by \cite{Ostrowski2020} examined lower bounds on circuit depth of QAOA. Their work identified $\text{MaxCut}$, MaxIndSet and a subset of Vertex Covering and SAT problems suitable for QAOA. They argued that problems such as the knapsack and travelling salesperson problems are not.

In this paper, we perform an instance space analysis for the first time on quantum algorithms, aiming to provide specific insights into the question of parameter concentrations of QAOA for $\text{MaxCut}$, but also opening the opportunity for ISA to be applied in a similar manner to explore many of the open questions faced by the quantum computing research community about how instance characteristics affect algorithm performance, and encouraging a greater diversity of instance classes to be explored than is current practice in the literature.

The structure of the remainder of this paper is as follows. In Section 2 we present the background necessary for applying QAOA to solve the $\text{MaxCut}$ problem, and discuss related experimental studies. In Section 3 we present the ISA framework, and the experimental settings to derive insights into how characteristics of $\text{MaxCut}$ instances impact the success of various QAOA initialisation strategies. The results are presented in Section 4 in the form of visualisations of the instance space and energy landscapes. We conclude the study with a discussion of key findings and future research directions in Section 5.

\section{QAOA for $\text{MaxCut}$}\label{background} 
\subsection{$\text{MaxCut}$}
Our case study for this paper focuses on the $\text{MaxCut}$ problem, which aims to find the  ``maximal cut" of a given graph, defined as follows:
\begin{definition}[$\text{MaxCut}$ Problem]
Given a simple undirected graph $G = (V, E)$, the objective of the $\text{MaxCut}$ problem is to find a partition of the vertex set $V$ into two subsets $S$ and $\bar{S}$ (where $\bar{S} = V \setminus S$) such that the number of edges in $E$ with one endpoint in $S$ and the other in $\bar{S}$ is maximised. Here, $V$ is the set of vertices $\{1, \dots, N\}$ and $E$ is the set of edges, $E = \{(i, j): i, j \in V, i \neq j\}$.
\end{definition}

\begin{figure}[H]
\begin{center}
\includegraphics[width=0.8\linewidth]{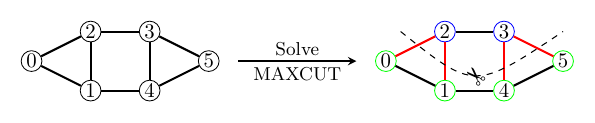}
\caption{An example of a six-node $\text{MaxCut}$ problem. The two partitions are defined as $S = \{1,2,5,6\}$ and $\bar{S} = \{3,4\}$.}
\label{fig:maxcut-example}
\end{center}
\end{figure}
To mathematically formulate the problem, we can express the objective function $C(\vec{z}) = \frac{1}{2} \sum_{(i,j) \in E} (1 - z_{i} z_{j})$, where $\vec{z}$ is a binary vector with $z_i = \pm 1$, indicating the partition of each vertex. Here, $z_i = 1$ if vertex $i$ is in set $S$ and $z_i = -1$ if it is in set $\bar{S}$. This formulation aims to maximise the number of edges between the two sets, which can also be rephrased as a minimisation problem:
\begin{align}
    \max C(\vec{z}) = \frac{1}{2} \sum_{(i,j) \in E} (1 - z_i z_j) \enspace \Rightarrow \enspace \min C(\vec{z}) = \frac{1}{2} \sum_{(i,j) \in E} (z_i z_j -1).
\label{eq:maxcut-obj}
\end{align}
The minimisation formulation can easily be represented as a QUBO formulation. We can introduce the transformation $z_i = 2x_i -1$ to rewrite the objective function in terms of binary variable $x_i$:
\begin{align}
    \min Q(x) = \sum_{(i,j) \in E} \frac{1}{2}((2x_i -1)(2x_j -1) - 1) = \sum_{(i,j) \in E} (x_i + x_j - 2x_i x_j)
\end{align}
In his ground-breaking paper, \cite{Karp1972} proved that $\text{MaxCut}$ can be reduced to the partition problem, an already demonstrated NP-hard problem. Assuming $P \neq NP$, NP-hard class problems are those for which no known polynomial time algorithm exists. Due to this, even most classical approaches to solving $\text{MaxCut}$ have been focused on finding near-optimal solutions. Most famously, \cite{Goemans1995} employ semi-definite programming (SDP) to achieve an approximation ratio of 0.878 with time complexity of $O(n^2 \log(n))$. This marked a significant advancement in approximation techniques and remains a gold standard. Many classical studies have improved this in recent years \citep{Fernandez2020}, \citep{Proença2023}, \citep{Bliznets2023}.
\\\\

\subsection{Quantum Approximate Optimisation Algorithm (QAOA)}
In this section, we will be discussing the Quantum Approximate Optimization Algorithm (QAOA). For those who may not be familiar with the bra-ket notation and quantum mechanics, we recommend referring to \cite{Nielsen2010} for a comprehensive introductory text on quantum computing. This source provides detailed explanations of these concepts. QAOA can be viewed as a particular case of a Variational Quantum Eigensolver (VQE) \citep{Peruzzo2014}. The VQE algorithm was initially designed to tackle quantum chemistry problems, aiming to approximate the ground state of a complicated Hamiltonian, a fundamental problem in physics and chemistry applications. The general premise of VQE algorithms is as follows:

\begin{enumerate}
    \item \textbf{Forming a Trial State}: We start by selecting a trial quantum state, denoted as $\ket{\psi(\vec{\theta})}$, as a function of parameters $\vec{\theta}$. This state is constructed using a parameterised quantum circuit $U(\vec{\theta})$ applied to an initial state $\ket{\psi(\vec{\theta})}$.
    \item \textbf{Computing the Expectation Value of the Hamiltonian}: The quantum computer is used to evaluate the expectation value of the Hamiltonian $H$ with respect to the trial state, represented as $\bra{\psi(\vec{\theta})}H\ket{\psi(\vec{\theta})}$.
    \item \textbf{Classical optimisation}: A classical optimiser adjusts the parameters $\vec{\theta}$ to minimise this expectation value. According to the variational principle, this expectation value provides an upper bound to the ground state energy of $H$, and ideally, minimising it leads us as close as possible to the actual ground state energy.
\end{enumerate}

QAOA was first introduced by \cite{Farhi2014}, and it can be seen as a special case of a VQE algorithm that solves combinatorial optimisation problems by minimising the energy of a problem Hamiltonian constructed using a QUBO formulation. As it lends itself to problems that can be represented by QUBOs and because of the ubiquity of real-world optimisation tasks being combinatorial optimisation problems \citep{Papadimitriou1998} which can be mapped to QUBOs, QAOA has garnered much interest in quantum computing research.  These optimisation problems are defined on $N$-bit strings $\vec{z} = (z_1, \dots, z_N)$ where the objective is to maximise an objective function $C(\vec{z})$.  In QAOA the objective function $C(z)$ is encoded into a Hamiltonian $H_C$ which is then minimised with respect to an ansatz state $\ket{\psi(\gamma, \beta)}$. This quantum state is evolved by unitary operators $H_B$, $H_C$ being repeated $p$ times:
\begin{align}
\ket{\psi_p(\gamma, \beta)} &= \prod_{j=1}^{p} e^{-i\beta_j H_B} e^{-i\gamma_j H_C} \ket{+}^{\otimes n}
\label{eq:ansatz}
\end{align}
where $\vec{\gamma} = (\gamma_1, \dots, \gamma_p)$ and $\vec{\beta} = (\beta_1, \dots, \beta_p)$ are the parameters being optimised, with $\gamma_i \in (-\pi, \pi)$ and $\beta_i \in (-\frac{\pi}{2}, \frac{\pi}{2})$. The $\ket{+}^{\otimes n}$ state is the ground state of $H_B = \sum^{n}_{j=1}X_j$ where $X_j$ is the Pauli $X$ operator acting on qubit $j$.

Based on our $\text{MaxCut}$ instance and its objective function in Equation (\ref{eq:maxcut-obj}), we can define $H_C$ as the cost Hamiltonian. For a given unweighted graph $G = (V, E)$ we can define the Hamiltonian as:
\begin{align}
    H_C = \frac{1}{2} \sum_{(j,k) \in E} (\mathbb{I} - Z_j Z_k)
\label{eq:cost-ham}
\end{align}
where $Z_j$ and $Z_k$ represent the Pauli $Z$ operator acting on qubit $j$ and $k$ respectively. The energy (or expectation) of the ansatz in Equation (\ref{eq:ansatz}) is given by:
\begin{align}\label{eq:energy-func}
F_p(\gamma, \beta) &= \braket{\psi_p(\gamma, \beta) | H_C | \psi_p(\gamma, \beta)}
\end{align}

\begin{figure}[H]
\centering
\includegraphics[width=\linewidth]{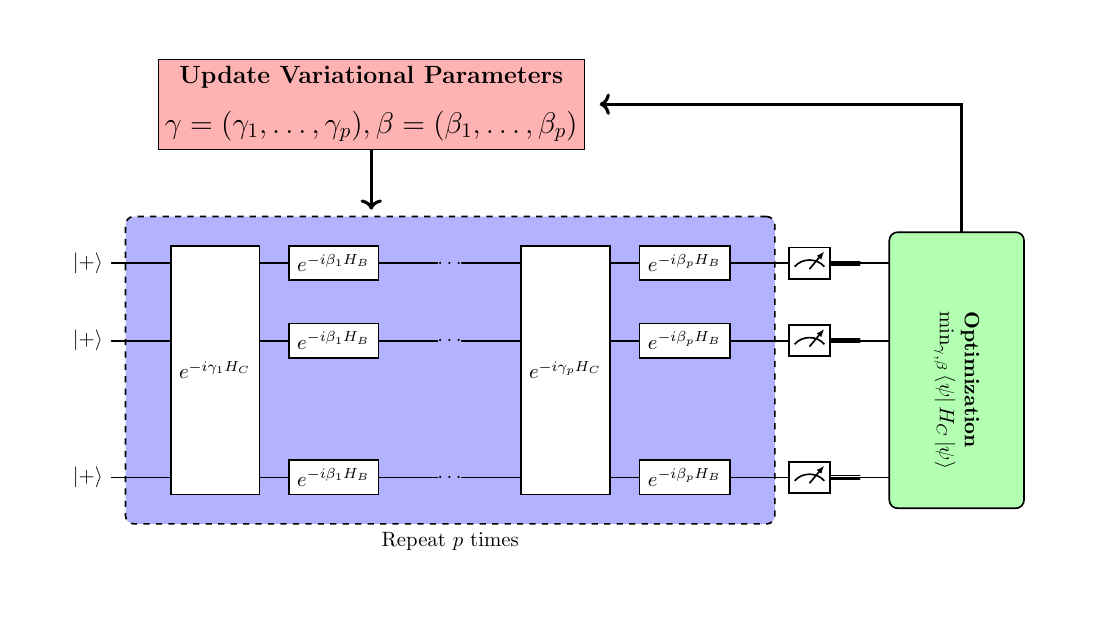}
\caption{Schematic of the QAOA ansatz: where each layer $i \in \{1, \dots, p\}$ consists of mixing unitary with parameters $\gamma_i$ and $\beta_i$ respectively.}
\label{fig:qaoa-schematic}
\end{figure}

As this expectation is parameterised by $\gamma$ and $\beta$, we can use a classical optimisation routine to search for angles that minimise $F_p$:
\begin{align}
(\gamma^*, \beta^*) &= \underset{\gamma,\beta}{\mathrm{arg\,min}}\ F_p(\gamma, \beta),
\end{align}
The diagram in Figure \ref{fig:qaoa-schematic} presents how this approach works. We initialise the parameters $(\gamma, \beta)$, and then construct a circuit based on the QAOA ansatz represented by Equation (\ref{eq:ansatz}), then based on the results of our measurement, we update $(\gamma, \beta)$ until the classical optimisation routine converges and optimal parameters $(\gamma^*, \beta^*)$ are obtained.

The success of the algorithm is given by its \textit{approximation ratio} $\alpha$ which is defined as:
\begin{align}
    \alpha := \frac{F_p(\gamma^*, \beta^*)}{C_{\max}}, \qquad \alpha \in [0,1]
\end{align}
where $C_{\max}$ is the maximum cut value for $G$. The approximation ratio is a widely used evaluation metric and indicates how near the solution given by QAOA is to the actual solution. The closer $\alpha$ is to 1, the closer we are to the optimal solution. As we increase the circuit depth $p$, $F_p(\gamma^*, \beta^*)$ will also increase, and as $p \rightarrow \infty$, $F_p(\gamma^*, \beta^*)$ approaches $C_{max}$ with $\alpha \rightarrow 1$ \citep{Farhi2014}.  Experimental results of QAOA have shown that at shallow depths, QAOA has worst-case performance guarantees that outperform random guessing, but not classical algorithms, for a variety of problems \citep{Farhi2014}, \citep{Halperin2002}.  However, finding a globally optimal $(\gamma^*, \beta^*)$ is not guaranteed. This is often due to the landscape of the cost function, which has been studied to suffer from barren plateaus \citep{Larocca2022}. This means that the landscape of the cost function can be flat, meaning that the gradients with respect to $(\gamma, \beta)$ vanish to be exponentially small with respect to the number of qubits. In fact, in their work \cite{Larocca2022}, they discuss how the input state can lead to the presence or absence of barren plateaus in QAOA.

\subsection{QAOA Optimisation Landscape}\label{sec2:qaoa-landscape}
The difficulty of finding optimal parameters $(\gamma^*, \beta^*)$ such that they minimise Equation (\ref{eq:cost-ham}) depends on the structure of the landscape. Several things can influence the structure of a landscape. In their work \cite{Munoz2017} identify several features of landscape characteristics that have been shown to predict the performance of algorithms. They also note that changes to landscape features can induce easy-hard phase transitions in algorithm performance.

For QAOA, the optimisation landscape can be easily visualised for $p=1$ as shown in Figure \ref{fig:qaoa-landscapes-2d} for different instance classes (see Appendix \ref{appendix:defs}). \cite{Lee2021} offers insights into how the QAOA landscape alters at higher dimensions. Their study observes for a QAOA ansatz with $p=2$, that when parameters $(\gamma_1, \beta_1)$ are fixed to minima, the landscape for $(\gamma_2, \beta_2)$ contains more low-value regions for $F_p(\gamma, \beta)$, and conversely, when $(\gamma_1, \beta_1)$ are fixed to maxima, the landscape consists of more high-value regions. This is extended to $p=10$, where they note that the maximum and minimum points slowly transform into lines. However, it is critical to note that their investigation is limited to 8-node 3-regular graphs, and other classes of instances are not examined in this analysis.

\begin{figure}[H]
\centering
\includegraphics[width=\linewidth]{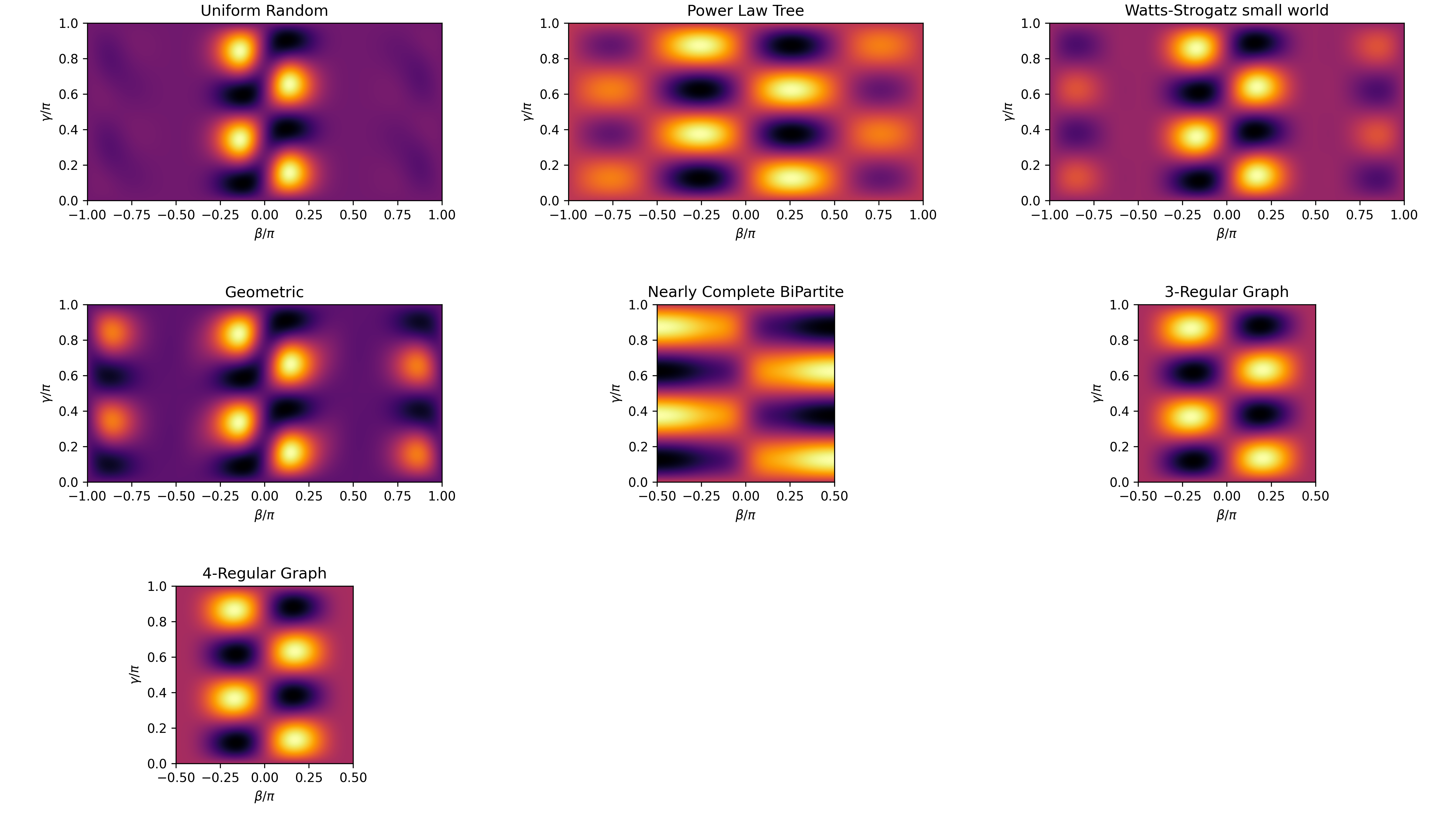}
\caption{QAOA landscapes across different instance classes for an ansatz of $p=1$ and system size $n=12$. The axes represent $\beta_1/\pi$, $\gamma_1/\pi$. The landscapes are symmetrical for each of the instance classes past these bounds.}
\label{fig:qaoa-landscapes-2d}
\end{figure}

To investigate how instance classes can influence the structure of a landscape, we generated $n=12$ node graphs for each class and then plotted the landscape at $p=1$. The visualisations of the QAOA landscapes as presented in Figure~\ref{fig:qaoa-landscapes-2d} underscore the necessity of considering the characteristics of individual instances when analyzing the performance of quantum algorithms. We can see that \textit{Geometric} graphs have a distinct landscape for where their minima exist. We also notice that both 3-regular and 4-regular graphs have very similar landscapes. The landscape for Power-Law Trees is also distinct with the presence of more local minima. These differences in the landscape for each graph instance suggest that a one-size-fits-all approach may not be optimal, as current studies suggest. Instead, a tailored \textit{instance dependent} strategy that accounts for each instance's unique features could yield more efficient and effective optimisation results. Although these visualisations only show the landscape for $p=1$. These landscapes will only get more distinct as we increase $p$.

\subsection{Parameter Initialisation}\label{sec2:qaoa-initialisation}

Many studies have shown that initial parameter settings for $(\gamma^*, \beta^*)$ impact the success of the QAOA \citep{Lee2021}, \citep{Zhou2020}, \citep{Scriva2023}. \cite{Sack2021} proposed an annealing-inspired strategy for parameter initialisation and \cite{Alam2020} used Machine Learning to improve parameter initialisation. \cite{Zhou2020} developed an approach specific to $\text{MaxCut}$. They utilise heuristic strategies such as INTERP and FOURIER in their work. INTERP involves using linear interpolation to extend optimal parameters from layer $p-1$ into $p$. FOURIER uses the discrete $\sin(\theta)$ and $\cos(\theta)$ transforms to determine $(\gamma^*, \beta^*)$ at $p$ from the optimal Fourier amplitudes found in $p-1$. \cite{Lee2021} proposed an approach for large-depth circuits in which the QAOA initialises layer $p$ with the optimal parameters from layer $p-1$. More recently, \cite{Sack2021} proposed a technique known as \textit{Trotterized Quantum Annealing} (TQA). This parameter initialisation method is inspired by the adiabatic time evolution of a quantum system \citep{Farhi2000}. \cite{Galda2021} proposed an instance-based approach. Their approach showed that parameters optimised for a small 6-node random graph can be applied to a 64-node random graph and achieve nearly the same performance with less than a 1\% decrease in $\alpha$. While this result indicates that initial parameters can be transferred from one instance to another, it is important to note that this work was limited to only $d-$regular and Erdos-Renyi graphs. \cite{Jain2022} introduced an initialisation method that utilises Graph Neural Networks (GNNs) as a warm-starting technique for QAOA. This approach enables generalisation over different graph sizes and thus can accelerate inference.  These highlighted works propose useful and novel ways to improve parameter initialisation.

\subsection{Related Work on Instance Characteristics}
To date, there has been little work in the literature exploring how instance characteristics across diverse classes of $\text{MaxCut}$ instances impact algorithm performance,  other than the work by \cite{Herrman2021} and \cite{Galda2021}. In Table \ref{table:paper-comparison}, we highlight some key studies on QAOA for $\text{MaxCut}$, we can see that most studies investigating QAOA's effectiveness in solving $\text{MaxCut}$ have focused on $d-$regular or Erdős–Rényi graphs.

\begin{table}[H]
\small 
\begin{tabularx}{\linewidth}{p{3cm} X X X X} 
\hline
\textbf{Reference (year)} & \textbf{Method} & \textbf{Initialisation} & \textbf{Instances} & \textbf{Year} \\ \hline
\cite{Farhi2014} & QAOA & & $d$-regular & 2014 \\ 
\cite{Brandao2018} & QAOA & Parameter-fixing & 3-regular & 2018 \\ 
\cite{Streif2019} & QAOA & Tree-QAOA (MPS) & 3-regular & 2019 \\
\cite{Zhou2020} & QAOA & INTERP, FOURIER & $d$-regular & 2019 \\
\cite{Moussa2020} & QAOA & & 4-regular & 2020 \\ 
\cite{Sack2021} & QAOA & Trotterized Quantum Annealing  & 3-regular, Erdos-Renyi & 2021 \\ 
\cite{Boulebnane2021} & QAOA & Analytical Guess & Erdos-Renyi, Chung-Lu & 2021 \\ 
\cite{Egger2021} & QAOA, Recursive QAOA & Warm-start & Fully-connected & 2021 \\ 
\cite{Marwaha2021} & QAOA & & d \textgreater{} 5 - regular & 2021 \\ 
\cite{Wurtz2021} & QAOA & & 3-regular & 2021 \\ 
\cite{Lee2021} & QAOA & Parameter-fixing & 3-regular, Erdos-Renyi & 2021 \\ 
\textbf{\cite{Herrman2021}} & \textbf{QAOA} & \textbf{} & \textbf{All connected non-isomorphic graphs} & \textbf{2021} \\ 
\cite{Galda2021} & QAOA & Parameter Transfer & $d$-regular, Erdos-Renyi & 2021 \\ 
\cite{Moussa2022} & QAOA, Recursive QAOA & Unsupervised ML & Erdos-Renyi, Chung-Lu & 2022 \\ 
\cite{Lee2022} & QAOA & Depth-progressive & 3-regular, 4-regular, Erdos-Renyi & 2022 \\ 
\cite{Jain2022} & QAOA & GNN & 3-regular & 2022 \\ \hline
\end{tabularx}
\caption{Table of recent studies on QAOA and the investigated instance classes. \cite{Herrman2021} highlighted as they investigated instance characteristics.}
\label{table:paper-comparison}
\end{table}

The paper by \cite{Herrman2021} studied how the structure of graphs affects the performance of QAOA. They used Pearson's correlation coefficient to analyze the impact of different graph features. Specifically, they investigated all the connected non-isomorphic graphs up to 8 nodes, which were generated according to \cite{Mckay2013}. The graph features studied include the number of edges, diameter, clique number, number of cut vertices, number of minimal odd cycles, group size, number of orbits, and the characteristics of being bipartite, Eulerian, or distance regular. QAOA performance was determined based on four metrics:

\begin{itemize}
  \item $C$: The expected value of $C$
  \item $P(C_{\text{max}})$: The probability of measuring a state that represents a maximum cut
  \item Level $p$ approximation ratio: $\frac{\langle C \rangle_p}{C_{\text{max}}}$
  \item Percent change in approximation ratio ($\Delta$ ratio) at $p$:
  $\frac{\langle C \rangle_p - \langle C \rangle_{p-1}}{1 - \frac{\langle C \rangle_{p-1}}{C_{\text{max}}}} = \frac{(C)_n - (C)_{n-1}}{C_{\text{max}} - (C)_{n-1}}$
\end{itemize}

The key findings identified some features, such as the presence of \textit{odd cycles} and the amount of symmetry in the graph, as strong predictors of QAOA success.  

Our work expands on this analysis in four significant ways. Firstly, we introduce a more diverse set of instances. Secondly, we study the impact of a more comprehensive set of features. Thirdly, we employ the ISA methodology as a means to understand the influence of instance characteristics (rather than solely using Pearson correlations). Lastly, we also examine how the features influence the performance of various initialisation techniques.

\section{Instance Space Analysis}\label{sec:methods}
\subsection{Overview of Instance Space Analysis}
Instance Space Analysis (ISA) is a recently developed methodology that supports the objective testing of algorithms and evaluates the diversity of test instances \citep{smith2023instance}. This approach draws inspiration from the algorithm selection framework presented in \cite{Rice1976} and the ``No Free Lunch Theorem" \citep{Wolpert1997}. The algorithm selection problem can be articulated as follows: 
\begin{definition}[Algorithm Selection Problem]
For a given $x \in \mathcal{P}$ with a feature vector $f(x) \in \mathcal{F}$, find a selection mapping $S(f(x))$ such that the selected algorithm $\alpha \in \mathcal{A}$ maximises performance $y(\alpha, x) \in \mathcal{P}$.    
\end{definition}

Rice's framework is highlighted in the blue region in Figure \ref{fig:instancespace}. In the framework by \cite{Rice1976}, for a given instance $x \in \mathcal{I}$ and algorithm $\alpha \in \mathcal{A}$, a feature vector $f(x) \in \mathcal{F}$ and performance metric $y(\alpha, x) \in \mathcal{Y}$ are measured. The process is repeated on all instances in $\mathcal{I}$ and algorithms in $\mathcal{A}$. This is used to generate $\{\mathcal{P}, \mathcal{F}, \mathcal{A}, \mathcal{Y}\}$, the  meta-data set for the problem. \cite{Rice1976} used  regression models on this meta-data to find a relationship between features and performance. 

ISA \citep{smith2023instance} extends the algorithm selection framework by \cite{Rice1976} by visualising the instance space and the performance of different algorithms in the instance space. The key steps of this framework are summarised in Figure \ref{fig:instancespace}. The vector spaces in the instance space analysis framework are consistent with those in the algorithm selection framework and are outlined as:
\begin{itemize}
    \item The problem space $\mathcal{P}$ represents the set of all (infinitely many) possible instances.
    \item The set of pre-existing or generated instances $\mathcal{I}$.
    \item The feature space $\mathcal{F}$ is a higher dimensional subspace that contains a vector representation of each problem instance defined by a set of features to characterise similarities and differences between instances.
    \item The algorithm space $\mathcal{A}$ is a set containing the collection of candidate algorithms.
    \item  The performance space, where $y \in \mathcal{Y}$ provides a comparable metric to compare performance across algorithms.
\end{itemize}

The two-dimensional projection of the feature space is constructed by using the meta-data to learn some mapping $g((f(x), y(\alpha, x))$ such that the variance of higher dimensional features is still captured and a kernel function is used to separate regions of poor performance from regions of good performance. This allows one to identify which algorithms are better suited for different classes of instances. To perform ISA we use the tools available at the Melbourne Algorithmn Test Instance Library with Data Analytics (MATILDA) \citep{Smith-Miles2020}. MATILDA allows us to:

\begin{figure}[H]
\centering
\includegraphics[width=0.8\linewidth]{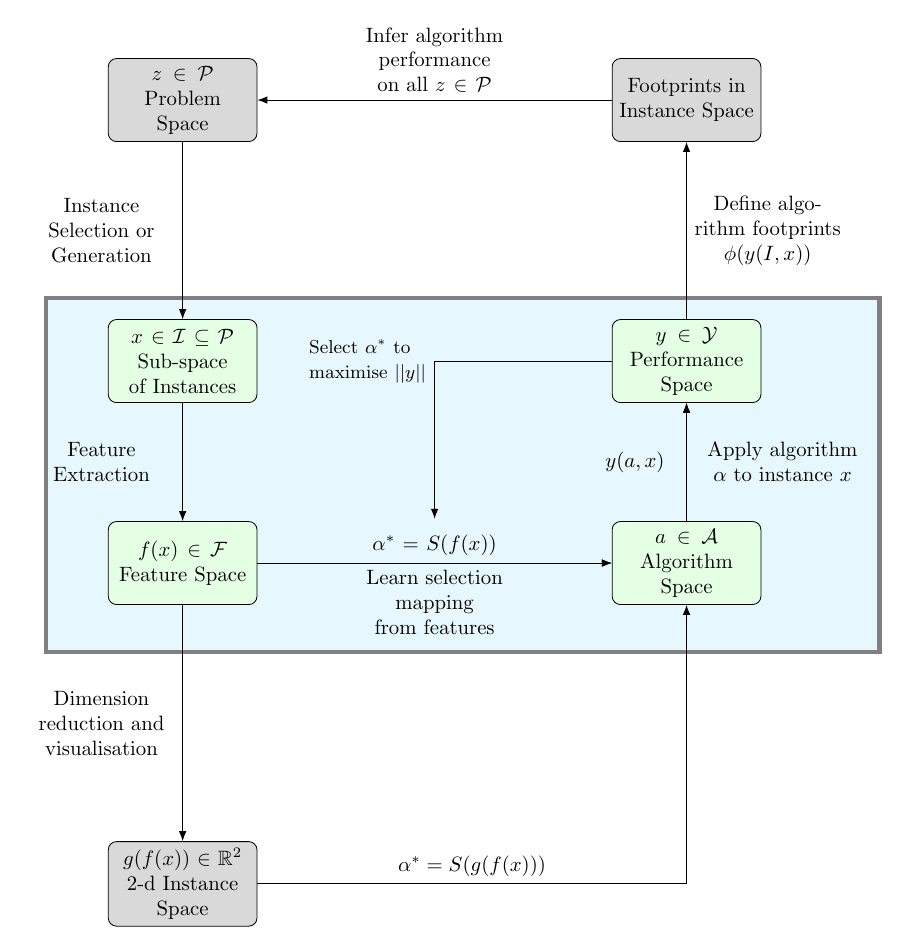}
\caption{A flowchart depicting the Instance Space Analysis framework \citep{smith2023instance}. Rice's algorithmic selection framework \citep{Rice1976} is highlighted in blue.}
\label{fig:instancespace}
\end{figure}

\begin{enumerate}
    \item Create a visualisation of the instance space for $\text{MaxCut}$ instances
    \begin{enumerate}
        \item Show the location of current benchmark instances across the instance space
        \item  Reveal the strengths and weaknesses of different parameter initialisation techniques
        \item Summarise the properties of the instances that an algorithm finds easy or difficult
    \end{enumerate}
    \item Determine objective measures of algorithmic success via footprint analysis
    \item Identify regions of the instance space where new test instances should be generated or acquired. This is especially important at the edges of the instance space to test the extreme cases, and at the edges of algorithm footprints to explore possible phase transitions
    \item Recommend the best algorithm for a given instance via automated algorithm selection.
\end{enumerate}

Once provided with meta-data, MATILDA performs feature selection to identify the most significant features that affect algorithm performance. MATILDA selects the most predictive features by computing cross-correlations as well as Pearson's correlation coefficient between features and algorithm performance. Based on these features, each instance is then represented by a high-dimensional feature vector. Based on user-defined criteria of acceptable performance, each instance is then labelled ``good" or ``bad" for each algorithm.  To select a subset of features for each algorithm $\alpha \in \mathcal{A}$, we first identify the features that correlate with performance. Similar features are clustered, and one feature per cluster is chosen. Machine learning techniques classify algorithm performance as ``good" or ``bad" for each cluster. The optimal feature set is the one that best separates ``good" or ``bad". Once this is done, MATILDA projects instances onto the 2D instance space using a method known as \textit{Projecting Instances with Linearly Observable Trends} (PILOT) \citep{Munoz2018}. Unlike Principal Component Analysis (PCA), which constructs principal components to capture the maximum variance in the data, PILOT adopts a different approach. PILOT minimises the following objective function:
\begin{align}
    \norm{\mathbf{F} - \mathbf{\hat{F}}} + \norm{\mathbf{y}^\intercal - \mathbf{\hat{y}}^\intercal}^2_{F} 
\end{align}
In this formulation, the matrices $\mathbf{F}$ and $\mathbf{y}$ represent the features and performance, respectively. The matrices $\mathbf{\hat{F}}$ and $\mathbf{\hat{y}}$ are their estimates. The objective of this function is to capture linear trends in the relationship between features and performance across the entire instance space. Additionally, this function ensures that the projection is predictive for both features and performance. The aim is to find two new axes $z_1$ and $z_2$, such that instances with similar features are located close to each other, and the direction of increasing and decreasing features are shown as clear trends in the plane. We can then easily visualise the algorithm performance across the instance space and see which instances have a ``good" or ``bad" performance for each algorithm. Next MATILDA uses the \textit{Correlated Limits of the Instance Space Theoretical or Experimental Regions} (CLOISTER) algorithm to generate the boundaries of the instance space. Once the boundaries of the instance space have been constructed, a heuristic algorithm (PYTHIA) is utilised to train a set of Support Vector Machines (SVMs) with either a Gaussian or Polynomial kernel based on each instance's ``good" or ``bad" labels. These SVM models are used for performance prediction and automated algorithm selection on unseen instances. The final step is to determine the algorithmic footprints. This is done using the Triangulation with Removal of Areas with Contradicting Evidence (\textit{TRACE}). An algorithms' footprint is defined as a region in the instance space where that algorithm is predicted to perform well. Footprint statistics and locations are used to provide an object assessment of each algorithm's strengths and weaknesses across the entire instance space. To understand the methodology in greater detail, we refer the reader to \cite{smith2023instance}. In the remaining subsections, we describe the meta-data for the ISA conducted on the MaxCut case study.

\subsection{Instance Space}\label{subsec:instance-classes}
The problem space $\mathcal{P}$, comprises all possible $\text{MaxCut}$. These are formally defined as follows. Let $G = (V, E)$. Where $V$ is the set of vertices $\{1, \dots N \}$ and $E$ is a set of edges, $E = \{ (i, j): i,j \in V, i \neq j\}$. Each graph instance can be generated from a different distribution. In this work, we focus on seven instance classes, namely: Uniform Random (Erdős–Rényi), Power-Law Tree, Watts-Strogatz Small World, Geometric, 3-regular, 4-regular, Nearly Complete Bipartite.
These classes were selected based on the diversity in features they present and their diverse applications in the real world.  In Figure \ref{fig:graphs} we present a collection of the instances classes examined in this work. See Appendix \ref{appendix:defs} for a more comprehensive set of definitions on how each instance class was constructed.

\begin{figure}[ht!]
  \centering
  \begin{subfigure}{0.3\textwidth}
    \includegraphics[width=\linewidth]{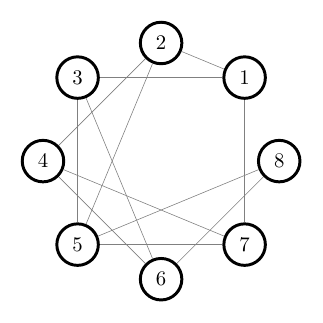}
    \caption*{(a)}
  \end{subfigure}
  \hfill
  \begin{subfigure}{0.3\textwidth}
    \includegraphics[width=\linewidth]{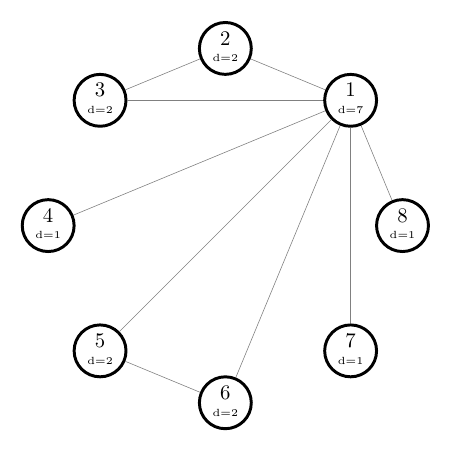}
    \caption*{(b)}
  \end{subfigure}
  \hfill
  \begin{subfigure}{0.3\textwidth}
    \includegraphics[width=\linewidth]{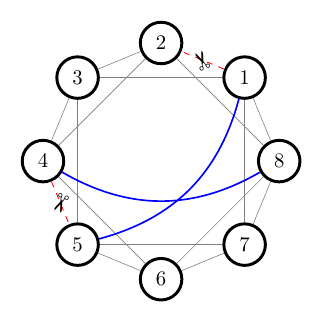}
    \caption*{(c)}
  \end{subfigure}

  \begin{subfigure}{0.3\textwidth}
    \includegraphics[width=\linewidth]{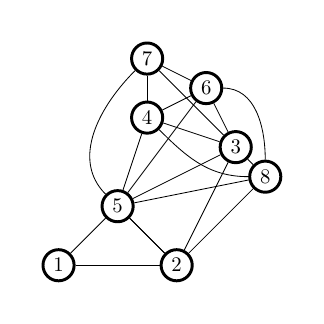}
    \caption*{(d)}
  \end{subfigure}
  \hfill
  \begin{subfigure}{0.3\textwidth}
    \includegraphics[width=\linewidth]{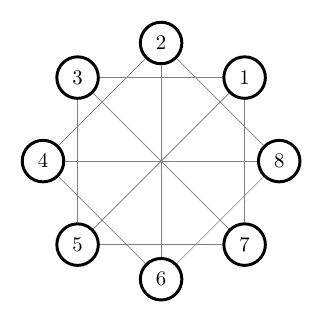}
    \caption*{(e)}
  \end{subfigure}
  \hfill
  \begin{subfigure}{0.3\textwidth}
    \includegraphics[width=\linewidth]{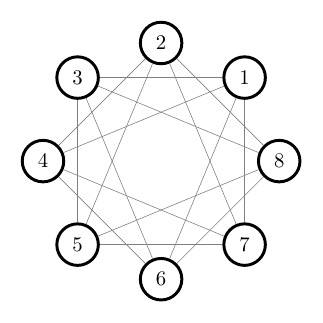}
    \caption*{(f)}
  \end{subfigure}

  \vspace{10pt} 
  \begin{subfigure}{0.3\textwidth}
    \caption*{} 
  \end{subfigure}
  \hfill
  \begin{subfigure}{0.3\textwidth}
    \includegraphics[width=\linewidth]{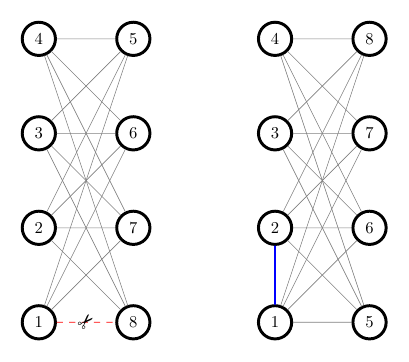}
    \caption*{(g)}
  \end{subfigure}
  \hfill
  \begin{subfigure}{0.3\textwidth}
    \caption*{} 
  \end{subfigure}

  \caption{Collection of graph diagrams: (a) Uniform Random (Erdős–Rényi), (b) Power Law Tree, (c) Watts-Strogatz Small World, (d) Geometric, (e) 3-Regular, (f) 4-Regular, (g) Nearly Complete Bipartite.}
  \label{fig:graphs}
\end{figure}
With a quick glance at Figure \ref{fig:graphs} one can quickly ascertain that each graph has a unique structure. \textit{Uniform Random (Erdős–Rényi)} graphs are where given a set of nodes $n$, every possible edge is included in the graph with a probability of $p$ \citep{Erdos1959}. \textit{Power-Law Trees}, have the property that the degree sequence of the graph follows a power-law distribution \citep{Albert2022}. These graphs have recently been used in applications such as modelling electricity grids, social e-commerce networks and trading patterns \citep{Somin2022}, \citep{Stephen2009} \citep{Hadaj2022}. \textit{Watts-Strogatz Small World} graphs are used to model the dynamics of ``small-world" networks \citep{Watts1998}.  These graphs model networks with both strong local clustering and short global separation, characterized by a high clustering coefficient and small average path length. \textit{Geometric graphs} are generated using an algorithm that places $n$ nodes in a unit cube, and then two nodes are connected by an edge if the distance between them is less than or equal to $r$ \citep{Penrose2003}. Figure \ref{fig:geometric-algorithm} in Appendix \ref{appendix:defs} shows an illustrative example of how this is done. These graphs are shown to have a community structure that is not generated by the existing instance classes. Both \textit{3-Regular} and \textit{4-Regular} graphs are examples of $d-$regular graphs. A regular graph with vertices of degree $d$ is called a $d$‑regular graph. These are the instances that have been studied in most quantum computing research thus far (see Table \ref{table:paper-comparison}). In the case of \textit{Nearly Complete Bipartite} graphs, we first construct a bipartite graph and then randomly add or remove an edge given some probability $p$. Such networks are often used in job matching \cite{Singh2023}.

\subsection{Feature Space}
It is crucial to ensure that the set of features $\mathcal{F}$ we measure on each instance are useful in order to make ISA effective. As per \cite{Smith-Miles2014}, useful features should possess two characteristics. Firstly, they should be computable in polynomial time, and secondly, they should expose characteristics that make an instance more challenging for a certain algorithm. We have identified 28 features to use for this first ISA for MaxCut, which are listed in Table \ref{tab:graph_features}. These features have been selected based on the research conducted by \cite{McAndrew2020}, \cite{Herrman2021} and also the work of \cite{Smith-Miles2014} in investigating graph colouring. For detailed definitions of each feature, please refer to Appendix \ref{appendix:feat-defs}. We compute a feature vector $f(x) \in \mathcal{F}$ for every instance. This vector contains all the features mentioned earlier for a single graph instance, and $\mathcal{F}$ is the feature space that encompasses all feature vectors for all graph instances in high dimension.

\begin{table}[h]
\centering
\scriptsize
\begin{tabularx}{\linewidth}{l X}
\hline
\textbf{Feature} & \textbf{Description} \\ 
\hline
\textbf{\textit{Structural Features}} &  \\ 
Number of Edges & Total count of edges in the graph. \\ 
Bipartite Graph & Indicates if the graph is bipartite (can be divided into two disjoint vertex sets). \\ 
Clique Number & Size of the largest complete subgraph (clique) in the graph. \\ 
Connected Graph & Indicates if the graph is connected (there is a path between every pair of vertices). \\
Density & A measure that reflects how many edges are in $G$ relative to the maximum possible number of edges among the vertices. \\
Edge Connectivity & The minimum number of edges whose removal makes the graph disconnected.\\ 
Maximum Degree & The highest degree among all vertices in the graph. \\ 
Minimum Degree & The smallest degree among all vertices in the graph. \\
Minimum Dominating Set & The smallest set of vertices such that every vertex is either in the set or adjacent to a vertex in the set. \\ 
Regular Graph & Indicates if the graph is regular (all vertices have the same degree). \\ 
Smallest Eigenvalue & The smallest eigenvalue of the graph's adjacency matrix. \\ 
Vertex Connectivity & The minimum number of vertices whose removal results in a disconnected or trivial graph. \\ 
\textbf{\textit{Cycle and Path based Features}} &  \\ 
Acyclic Graph & Indicates if the graph has no cycles. \\ 
Average Distance & The average length of the shortest paths between all pairs of vertices. \\ 
Diameter & The longest shortest path between any two vertices in the graph. \\ 
Eulerian Graph & Indicates if the graph has an Eulerian path or circuit. \\ 
Number of Components & The total number of connected components in the graph. \\ 
Planar Graph & Indicates if the graph can be drawn on a plane without edge crossings. \\ 
Radius & The minimum eccentricity of any vertex in the graph. \\ 
\textbf{\textit{Spectral Features}} &  \\ 
Algebraic Connectivity & The second-smallest eigenvalue of the Laplacian matrix, indicating graph connectivity strength. \\
Laplacian Largest Eigenvalue & The largest eigenvalue of the graph's Laplacian matrix. \\ 
Ratio of Two Largest Laplacian Eigenvalues & The ratio between the largest and second-largest eigenvalues of the Laplacian matrix. \\ 
Ratio of Two Smallest Laplacian Eigenvalues & The ratio between the smallest and second-smallest eigenvalues of the Laplacian matrix. \\ 
\textbf{\textit{Additional Features studied by \cite{Herrman2021}}} &  \\ 
Distance-Regular Graph  & Whether or not the graph is Distance Regular. \\ 
Group Size & The group size of $G$ is the size of the automorphism group of $G$. \\
Number of Cut Vertices & A cut vertex of a connected graph is a vertex whose removal disconnects the graph.\\ 
Number of Minimal Odd Cycles & Count of the smallest-sized cycles with an odd number of edges. \\ 
Number of Orbits  & Count of unique sets of vertices, where each set, or orbit, is formed by interchangeable vertices under the graph's automorphisms. 
\\ \hline
\end{tabularx}
\caption{Set of 28 features used in the metadata. For each graph instance $x$, we construct $f(x)$ based on the definitions provided in this table. See Appendix \ref{appendix:feat-defs} for a more comprehensive set of definitions.}
\label{tab:graph_features}
\end{table}

\subsection{Algorithm Space}
For the purposes of our ISA, we investigate four initialisation strategies for the QAOA ansatz at $p=3$ layers. As such, we will be investigating the initialisation of six parameters, namely $\gamma = (\gamma_1, \gamma_2, \gamma_3)$ and $\beta = (\beta_1, \beta_2, \beta_3)$, and each ``algorithm" corresponds to an initialisation strategy. The algorithms are as follows:
\begin{enumerate}
    \item Random Initialisation
    \item Trotterised Quantum Annealing (TQA)
    \item Instance-based Parameter Initialisation
    \item 3-regular Graph Initialisation
\end{enumerate}
Random initialisation simply involves sampling $\gamma_i, \beta_i$ from a uniform random distribution of $\gamma_i \in (-\pi, \pi)$ and $\beta_i \in (-\frac{\pi}{2}, \frac{\pi}{2})$ respectively. The TQA algorithm was first presented by \cite{Sack2021} as an approach to initialising QAOA parameters based on the Trotterized quantum annealing method (TQA); TQA is parameterised by the \textit{Trotter time step}. The algorithm is discussed in detail in \cite{Sack2021}. We now introduce a new approach to initialising parameters called \textit{Quantum Instance-based Parameter Initialisation} (QIBPI), or simply, Instance-based Parameter Initialisation.
\subsubsection{Instance-based Parameter Initialisation}
\begin{figure}[ht]
\centering
\includegraphics[width=\textwidth]{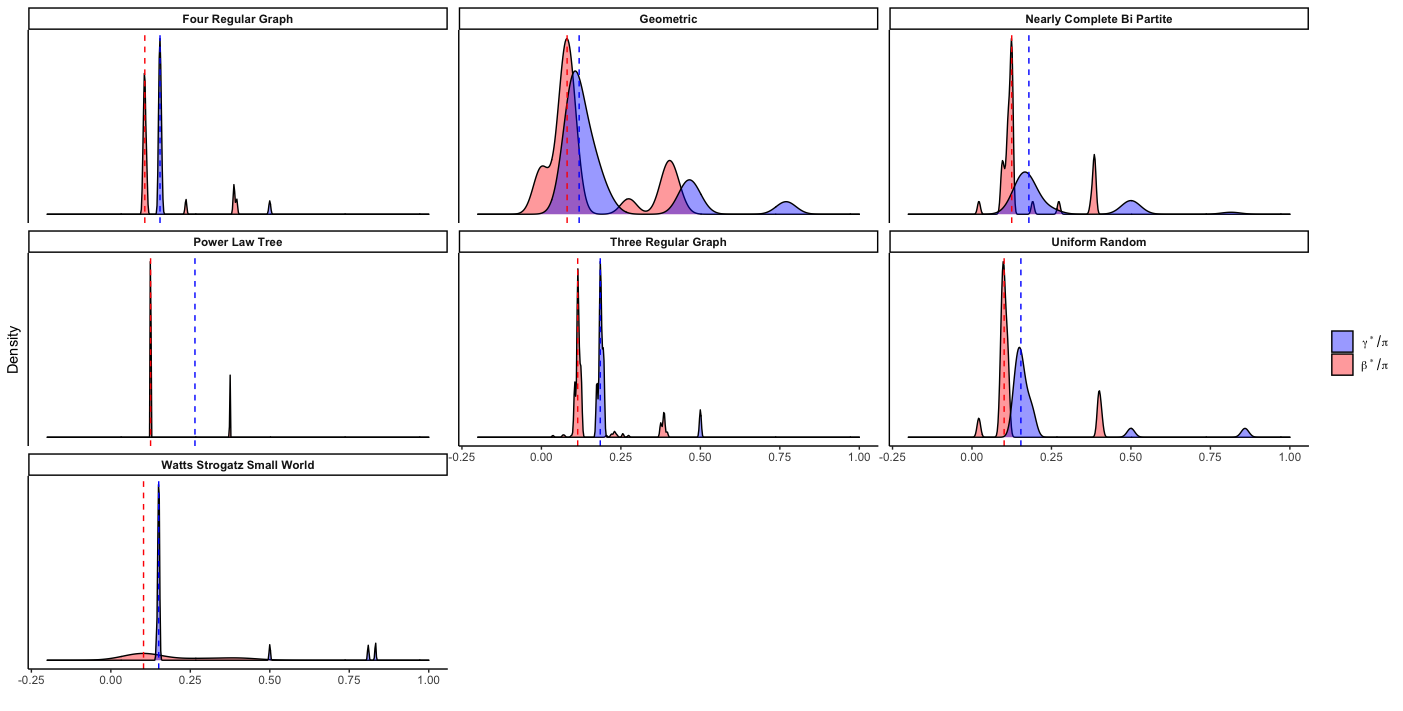}
\caption{Distribution of optimal parameters $(\gamma^*, \beta^*)$ by instance class for $p=1$. The dashed lines indicate the median value for each of these parameters.}
\label{fig:histogram-of-optimal-parameters-by-instance-class}
\end{figure}

Based on the insights in subsection \ref{sec2:qaoa-landscape}, we hypothesize that the efficacy of a given parameter initialisation strategy for QAOA could rely on the structure of the graph instances it operates upon. We propose a new approach called \textit{Quantum Instance-based Parameter Initialisation} (QIBPI) to tackle this. In short, our strategy involves constructing a diverse set of instances (the instance classes highlighted in subsection \ref{subsec:instance-classes}), then subsequently determining optimal parameters through simulation for each instance class. Then, during QAOA initialisation for a larger $n$, we utilise the optimal parameters of the given instance class as our initial parameters. This is a similar approach to that taken by \cite{Galda2021}. The first stage of our methodology is encapsulated in Algorithm \ref{alg:graphgen}, which is dedicated to identifying the optimal parameter distribution across each instance class. We generate 100 graph instances for each class. Then, for every graph instance $G$, QAOA is run with random initialisation, and we extract the optimal parameters $(\gamma^*_G, \beta^*_G)$. We then compile these parameters across all instances within the same class and calculate their median parameters for each layer, up to $L=10$ layers. These median optimal parameters are stored and categorized by graph class to facilitate efficient parameter initialisation for future QAOA runs on similar instances. The rationale behind this median-based approach is substantiated by the distribution depicted in Figure \ref{fig:histogram-of-optimal-parameters-by-instance-class}. We can see that each instance class has a narrow of distribution (exhibiting the parameter concentration phenomena). However, for ``Power Law Tree" instances, it appears simply using the median may not yield good performance as the distribution depicted is wider. We present the medians for $p=3$ in Table \ref{table:optimal-vals}. We have also pre-computed the median up to $p=10$ layers, and these values are available in the GitHub repository associated with this project \cite{Katial2023}.
\begin{table}[htp]
\centering
\small 
\begin{tabular}{lcccccc}
\hline
\textbf{Instance Class}    & $\gamma_1$ & $\gamma_2$ & $\gamma_3$ & $\beta_1$ & $\beta_2$ & $\beta_3$ \\ \hline
3-Regular Graph            & -0.1165                        & 0.2672                         & 0.1837                         & 0.3958                         & 0.2506                         & -0.1447                        \\
4-Regular Graph            & 0.1370                         & 0.0331                         & 0.1837                         & 0.3238                         & -0.6131                        & -0.2158                        \\
Uniform Random             & 0.1428                         & -0.2116                        & -0.1327                        & 0.2952                         & 0.1680                         & 0.1992                         \\
Geometric                  & -0.1564                        & 0.0487                         & -0.1783                        & 0.3514                         & -0.0762                        & 0.1305                         \\
Nearly Complete Bipartite  & 0.0958                         & -0.0715                        & -0.2107                        & 0.2473                         & -0.1765                        & 0.1576                         \\
Power Law Tree             & 0.2426                         & 0.2190                         & 0.1971                         & -0.5864                        & -0.3827                        & 0.2193                         \\
Watts Strogatz Small World & 0.1490                         & 0.2042                         & -0.1868                        & 0.3097                         & 0.2401                         & 0.1078                         \\ \hline
\end{tabular}
\caption{Table of median values for the optimal $(\gamma_p, \beta_p)$ for $p=3$. These are the median values across 100 graphs of each instance class.}
\label{table:optimal-vals}
\end{table}

\begin{algorithm}[htp]
 \scriptsize
\DontPrintSemicolon
\caption{Generate Optimal Parameter Distribution for QAOA}
\KwIn{Number of nodes $N = 8$, Number of layers $L=10$}
\KwOut{Median optimal parameters $\gamma$ and $\beta$ for each class}

\SetKwFunction{FMain}{GenerateGraphInstances}
\SetKwFunction{Foptimise}{optimiseQAOAParameters}
\SetKwFunction{FMedian}{CalculateMedianParameters}
\SetKwProg{Fn}{Procedure}{:}{\KwRet}
\Fn{\FMain{}}{
    \For{$p \leftarrow 1$ \KwTo $L$}{
        \ForEach{graph type $T$ in \{Uniform Random, Power-Law Tree, Watts-Strogatz Small World, Geometric, 3-regular, 4-regular, Nearly Complete Bipartite\}}{
            \For{$i \leftarrow 1$ \KwTo $100$}{
                Generate graph instance $G_{i, p, T}$ with $N$ nodes and layer $p$\;
                \Foptimise{$G_{i, p, T}$}\;
            }
        \FMedian{$T_p$}\;
        }
    }
}

\SetKwProg{Fn}{Function}{:}{\KwRet}
\Fn{\Foptimise{$G$}}{
    \KwIn{Graph instance $G$}
    \KwOut{Optimal parameters $\gamma_G$, $\beta_G$}
    Run QAOA on graph instance $G$ with random initialisation to find optimal parameters\;
    \KwRet{$\gamma_G$, $\beta_G$}\;
}

\SetKwProg{Fn}{Function}{:}{\KwRet}
\Fn{\FMedian{$T$}}{
    \KwIn{Graph type $T$, Set of parameters $\{\gamma_{i, p, T}\}$, $\{\beta_{i, p, T}\}$}
    \KwOut{Median parameters $\gamma_{median, T}$, $\beta_{median, T}$}
    Calculate the median of the parameters $\gamma_{i, p, T}$ and $\beta_{i, p, T}$ for all instances of type $T$\;
    Store median parameters in a parameter class corresponding to $T$\;
    \KwRet{$\gamma_{median, T}$, $\beta_{median, T}$}\;
}
\label{alg:graphgen}
\end{algorithm}

Our last algorithm involves initialising using $(\gamma, \beta)$ a 3-regular graph optimised parameters. We achieve this by re-using the experimentally computed optimal parameters for 3-regular graphs and applying them to other instance classes. The goal is to demonstrate how effective it can be to reuse parameters for 3-regular graphs on other instance classes. We denote this algorithm as ``3-Regular Graph Optimised" initialisation. Using these initial parameters will help us rigorously test the claim of whether or not optimal $(\gamma, \beta)$ parameters are in fact instance independent of the class of instances too. This gives us the following four algorithms in our portfolio: \textit{Random Initialisation}, \textit{Trotterised Quantum Annealing}, \textit{3-Regular Graph Optimised} and \textit{Instance Class Optimised}

\subsection{Performance Space}\label{sec:isa-performance-space}

\begin{figure}[htp]
    \centering
    \includegraphics[width=0.8\textwidth]{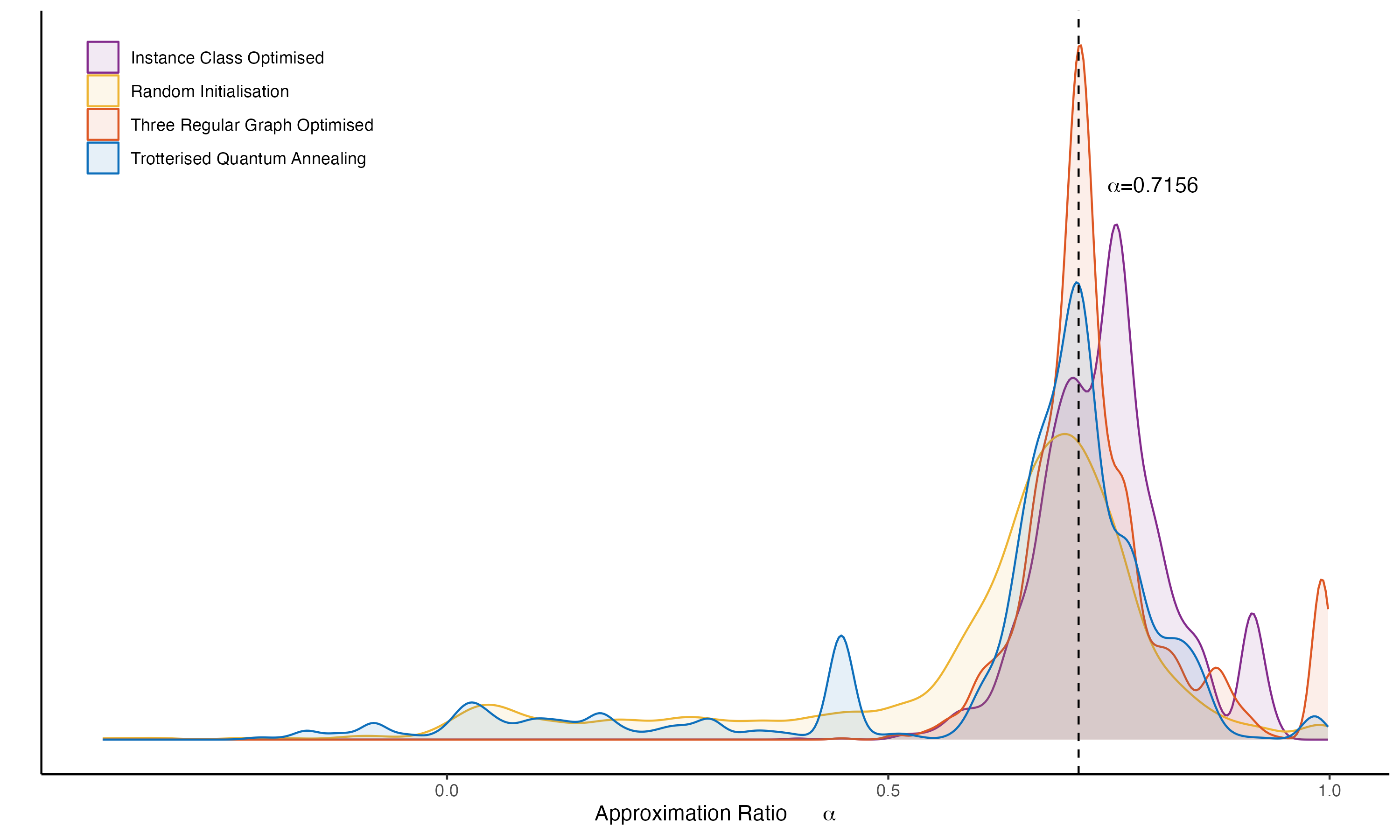}
    \caption{Histogram of the final approximation ratio of each initialisation strategy for a QAOA of $p=3$ and system size of $N=12$ nodes. The median approximation ratio of $\alpha = 0.7156$. This is across 2,100 instances of all classes.}
    \label{fig:histogram-of-optimal-parameters-by-initialisation}
\end{figure}

Typically, one might assess these methods using a standard metric such as the approximation ratio $\alpha$. However, as the data presented in Figure \ref{fig:histogram-of-optimal-parameters-by-initialisation} shows, over an extended period, most algorithms tend to reach a reasonably good approximation ratio, and the final approximation ratio concentrates to $ \alpha \approx 0.72$. This insight led us to develop a more nuanced metric for comparing algorithm performance. Our metric combines the number of function evaluations and the approximation ratio.

To apply this new metric: For each instance, we first determine the highest approximation ratio achieved by any initialisation algorithm, which we denote as $\alpha_{\max}$. We then define what we consider to be an acceptable approximation ratio, $\alpha_{\text{acceptable}}$, as 95\% (our threshold value $\tau$) of this maximum ratio. The performance of each initialisation algorithm, $\kappa$, is then assessed based on how many iterations are required to reach this acceptable approximation ratio. If a method does not achieve the acceptable level, we assign it a penalty score, set high at $M=1 \times 10^5$.

\begin{algorithm}[htp]
\scriptsize
\DontPrintSemicolon
\KwIn{A set of initialisation algorithms, each evaluated over multiple instances}
\KwOut{Efficiency scores for each initialisation algorithm}

\SetKwFunction{FMain}{EvaluateEfficiency}
\SetKwProg{Fn}{Function}{:}{}
\Fn{\FMain{Algorithms}}{
    $\alpha_{\max} \gets 0$\;
    \ForEach{instance in dataset}{
        \ForEach{algorithm in Algorithms}{
            $\alpha \gets$ ComputeApproximationRatio(algorithm, instance)\;
            \If{$\alpha > \alpha_{\max}$}{
                $\alpha_{\max} \gets \alpha$\;
            }
        }
    }
    $\alpha_{\text{acceptable}} \gets 0.95 \times \alpha_{\max}$\;
    \ForEach{algorithm in Algorithms}{
        functionEvaluations $\gets$ NumberOfEvaluationsToReach(algorithm, $\alpha_{\text{acceptable}}$)\;
        \eIf{functionEvaluations exists}{
            Score[algorithm] $\gets$ functionEvaluations\;
        }{
            Score[algorithm] $\gets M = 1 \times 10^5$\;
        }
    }
    \Return{Score}\;
}
\caption{Evaluation of Initialisation Algorithms Efficiency}
\end{algorithm}

\begin{figure}[htp]
    \centering
    \includegraphics[width=0.8\textwidth]{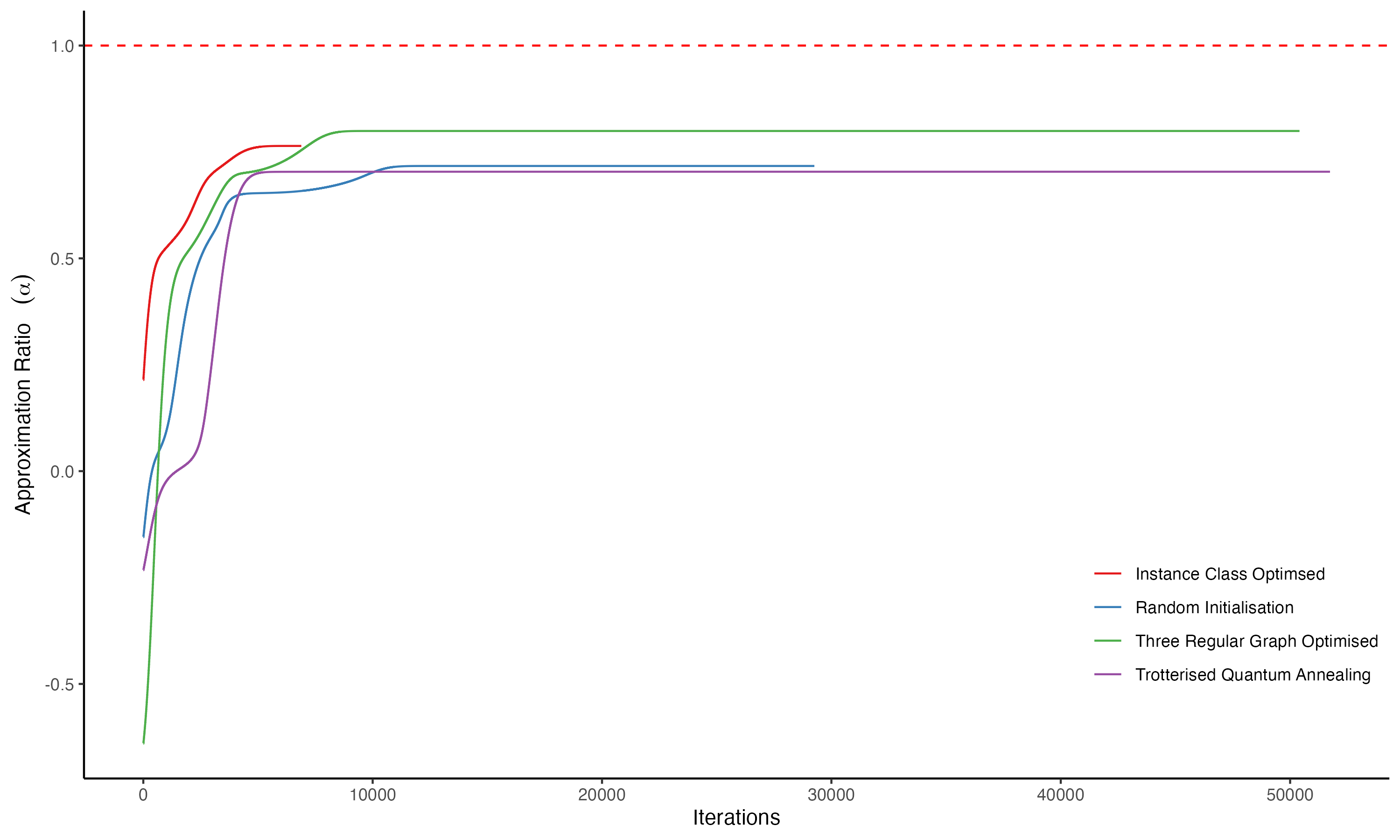}
    \caption{Approximation ratio of each initialisation strategy for a QAOA of $p=3$ and system size of $N=12$ nodes.}
    \label{fig:algo-perf-on-single-instance}
\end{figure}

The general method is now applied to a specific instance involving a ``4-regular" graph with 12 nodes.  We can see the performance of each initialisation algorithm in Figure \ref{fig:algo-perf-on-single-instance}.
Across the four algorithms (instance-based, 3-regular, TQA and random), we determine that the best performing algorithm was the ``3-regular" initialisation, achieving $\alpha = 0.7992693$. As such, $a_{\text{acceptable}}=0.95 \times \alpha_{\max} = 0.7593058$. The performance metric $\kappa$ is then determined for each algorithm based on the number of iterations required to reach or exceed the performance threshold. The performance metric for each algorithm in this case is highlighted in Table \ref{tab:performance_metrics}.
\begin{table}[h]
\centering
\small
\begin{tabular}{l c c c r}
\hline
\textbf{Algorithm} & \textbf{Iterations} & \textbf{$\alpha^*$} & \textbf{Performance Metric} ($\kappa$) & \textbf{Binary Label}\\
\hline
Instance Class Optimised & 6,889 & 0.7641341 & 4,796 &  ``good" \\
3-regular Graph Optimised & 50,409 & 0.7992693 & 7,029 & ``bad"\\
Random Initialisation & 29,258 & 0.7170581 & 100,000 (Penalty Value) & ``bad" \\
Trotterised Quantum Annealing &  51,739 & 0.7035441 & 100,000 (Penalty Value)& ``bad" \\
\hline
\end{tabular}
\caption{Performance Metrics for Quantum Optimisation Algorithms for an example ``4-regular" graph with 12 nodes, $\alpha^*$  is the final approximation ratio for each algorithm. An acceptable approximation ratio $\alpha_{\text{acceptable}}$ is 0.7593058. The binary label is based on whether or not each algorithm is within 10\% of the best performing algorithm.}
\label{tab:performance_metrics}
\end{table}
In our ISA we will also compare the algorithms based on a binary ``good" or ``bad" label. We will assign the performance of an algorithm to be ``good" if its performance metric $\kappa$ is within $100 \times \epsilon$ percent of the best performing algorithm -- where $\epsilon$ is a \textit{goodness threshold}. For our ISA we will use $\epsilon = 0.1$. In this example, the best $\kappa$ value was 4,796 iterations which was for the \textit{Instance Class Optimised} algorithms. As all other algorithms fail to fall within 10\% of this $\kappa$, the remaining algorithms are assigned a binary label of ``bad".

\section{An Instance Space for QAOA Initialisation Strategies on $\text{MaxCut}$}
Having completed the construction of our meta-data $\{\mathcal{I},\mathcal{F}, \mathcal{A}, \mathcal{Y} \}$, we can now use MATILDA to construct our instance space.

\subsection{Computational Settings}
To build our ISA, we leveraged the University of Melbourne HPC cluster, SPARTAN. Each simulation was conducted using the \texttt{qiskit} QASM Simulator \citep{Qiskit}. Within each simulation, we generate a single instance and initialise using each of the four techniques highlighted in Section \ref{sec2:qaoa-initialisation}. The classical optimiser utilised was ADAM \citep{Kingma2017}, and then after convergence, we conducted post-processing to compute our performance metrics. Each run was logged to an MLFlow server -- a framework we utilised to store, track and version our experiments \citep{Zaharia2020}. With this approach, we created 2,100 instances for our ISA by generating 300 graph instances of each of the 7 instance classes, with $N=12$ nodes in each instance. To generate these instances, we used Python's NetworkX library \citep{Hagberg2008}. The instance code is available on the GitHub repository \citep{Katial2023}. We used \texttt{ez-experimentr} to manage workloads \citep{Katial2020} and MATILDA to analyze the instance space after simulation completion.

\subsection{Instance Space}
As explained in Section \ref{subsec:instance-classes}, MATILDA produces a subset of features that have the most significant impact on algorithmic performance. For our study, the following linear transformation represents the projection matrix of selected features onto the 2D instance space:
\begin{align}
\small
\mathbf{Z} =
\begin{pmatrix}
    Z_1 \\
    Z_2 
\end{pmatrix}
        &= 
 \left(\begin{array}{@{}c@{\hspace{2em}}c@{}}
        0.5051 & -0.485  \\
        -0.6291 & 0.0463 \\
        0.4771 & -0.0263 \\
        -0.4878 & -0.9917 \\
        0.5781 & -0.0577 \\
        0.4284 & -0.2866 \\
        -0.0279 & 0.9336 \\
        -0.5347 & -0.4114 \\
        0.4849 & 0.9991 \\
        -0.4417 & 0.5989
    \end{array}\right)^{\intercal}
    \begin{pmatrix}
        \text{Radius} \\
        \text{Minimum Degree} \\
        \text{Minimum Dominating Set} \\
        \text{Regular} \\
        \text{Planar} \\
        \text{Average Distance} \\
        \text{Laplacian Largest Eigenvalue} \\
        \text{Number of Orbits} \\
        \text{Group Size} \\
        \text{Number of Edges}
    \end{pmatrix} 
\label{eq:projection-transformation}
\end{align}

\begin{figure}[H]
    \centering
    \includegraphics[width=\textwidth]{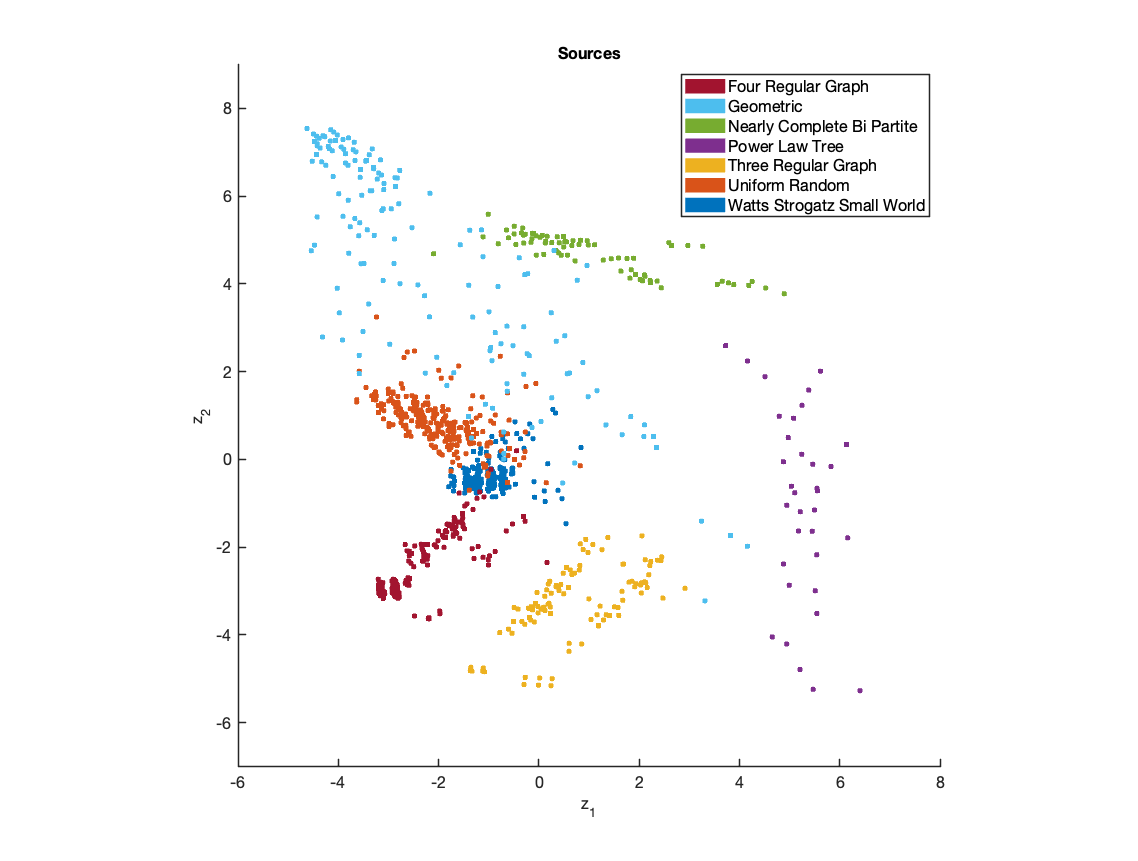}
    \caption{Distribution of the set of 2,100 12-node graphs across the Instance Space defined by Equation (\ref{eq:projection-transformation}), coloured by instance class.}
    \label{fig:source-dist}
\end{figure}

Equation \ref{eq:projection-transformation} shows how a particular instance, with a specific feature vector, is mapped as a point in the 2D instance space. The coordinates of the instance space are represented by $Z_1$ and $Z_2$. In Figure \ref{fig:source-dist}, we can observe the distribution of our seven instance classes across the instance space. Each point on the plot indicates a unique 12-node graph from the generated set of 2,100 graphs.

After analysing the distribution of all instances based on their class, it is evident that each instance class is located in a unique part of the instance space due to different features characterising those instances. It is notable that the density and spread along $Z_1$ and $Z_2$ varies significantly between instance classes. For example, the ``4-Regular Graphs" exhibit a tightly clustered group, which indicates low variance in our feature vector. In other words, it is very difficult for a diverse set of 4-regular graphs to be generated. On the other hand, ``Geometric" graphs exhibit higher variance and are represented by a much more dispersed cluster.  Some clusters are closer or even overlapping, such as ``3-Regular Graphs" and ``Watts-Strogatz" and some ``Uniform Random" graphs. It is not surprising that the Watts-Strogatz graphs are closely clustered together. This proximity indicates that the characteristics measured by our feature vector are similar, as expected, since the Watts-Strogatz random generation process produces graphs with comparable average clustering, average path length, and average node degree. This suggests that the generated graphs may be alike and share many of the features we have measured, which is why they are clustered together on the plane. However, we should keep in mind that our study focuses only on graphs of size $N=12$, which limits our ability to measure accurately the potential diversity that can be generated by different instance generation processes. Additionally, we have observed that the ``Power Law Tree" and ``Nearly Complete BiPartite" graphs are situated in distinct regions of the instance space as compared to other graphs. The most interesting insight here is that literature has focussed primarily on ``3-Regular Graphs" and ``Uniform Random" graph instances, which occupy a small part of the broader instance space, and therefore the conclusions of many studies may not generalise to other types of graphs.

\begin{figure}[ht]
\centering
\includegraphics[width=.22\linewidth]{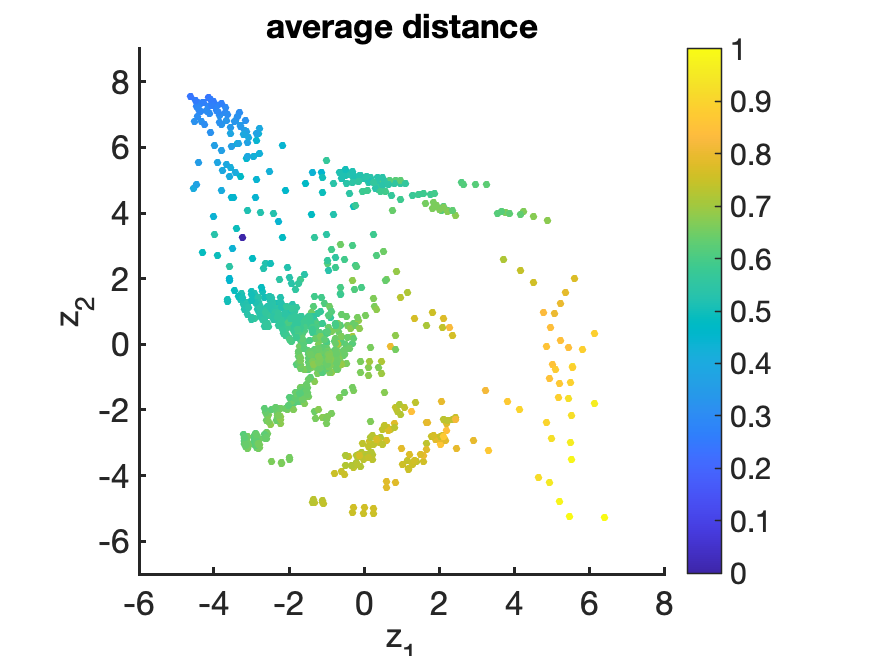}
\includegraphics[width=.22\linewidth]{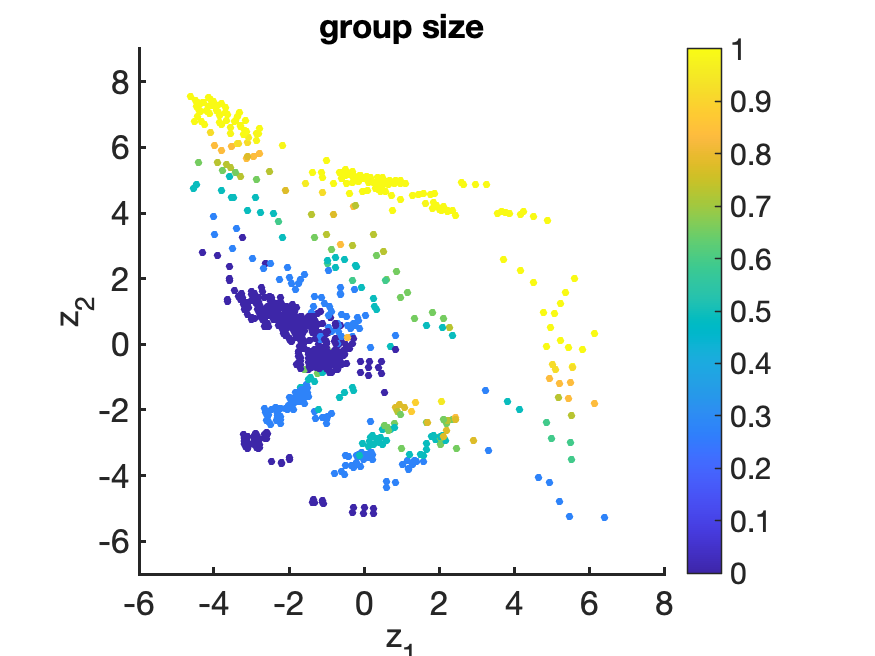}
\includegraphics[width=.22\linewidth]{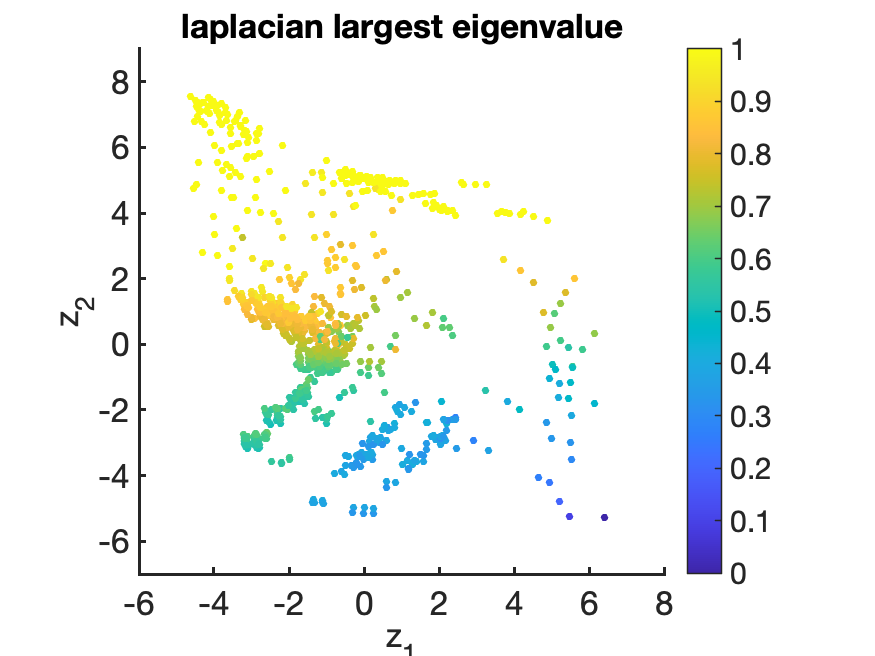}
\includegraphics[width=.22\linewidth]{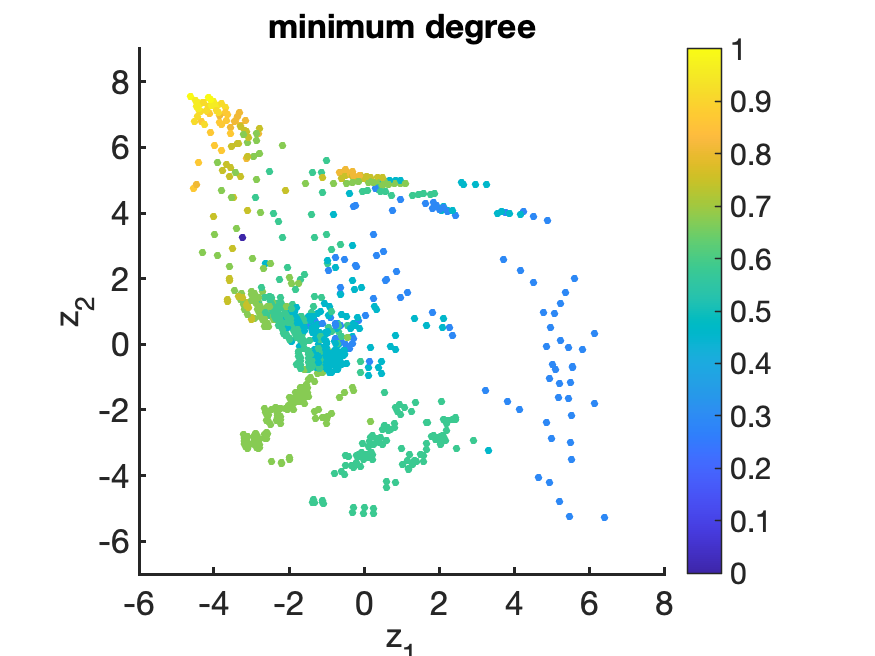}

\includegraphics[width=.22\linewidth]{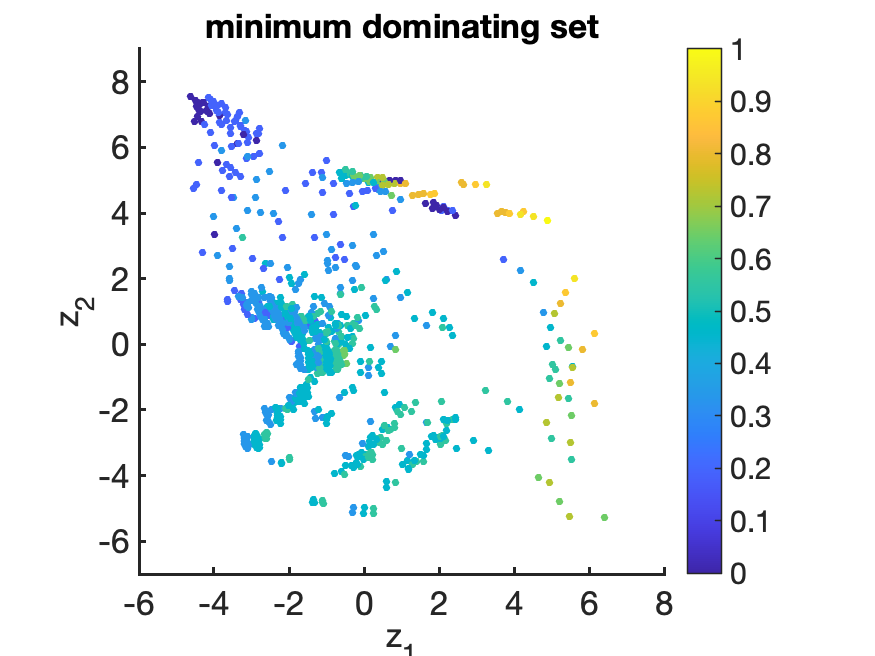}
\includegraphics[width=.22\linewidth]{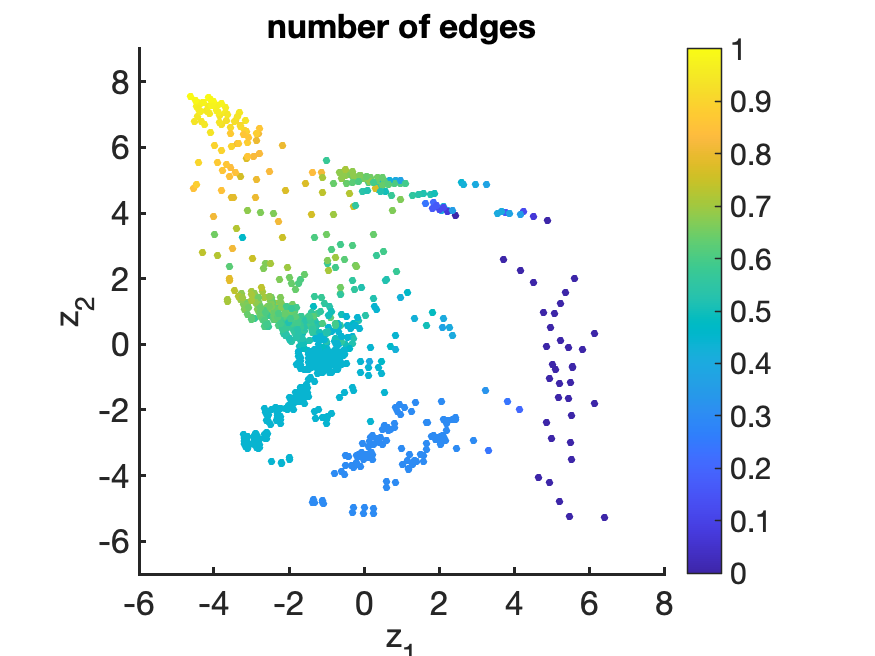}
\includegraphics[width=.22\linewidth]{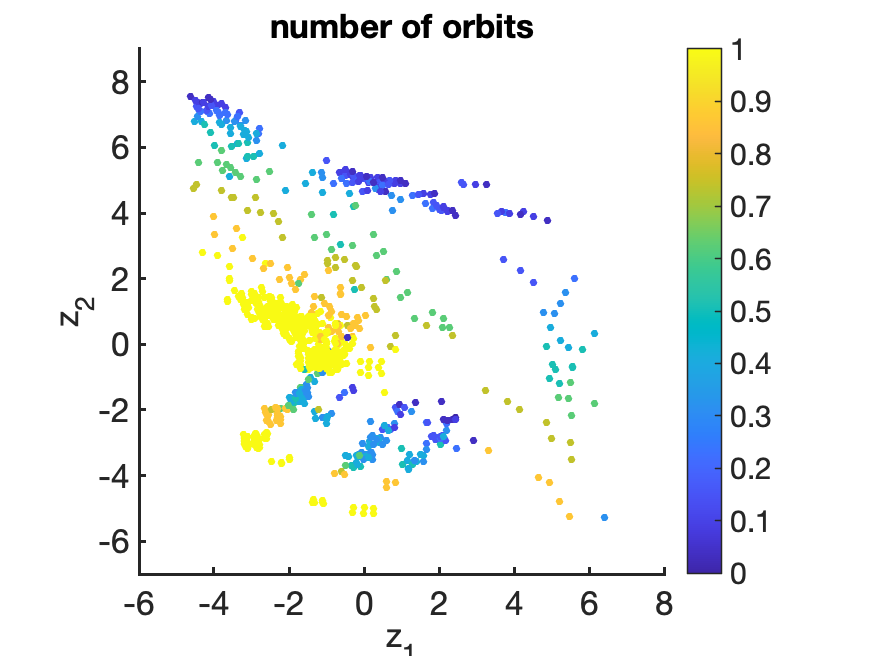}
\includegraphics[width=.22\linewidth]{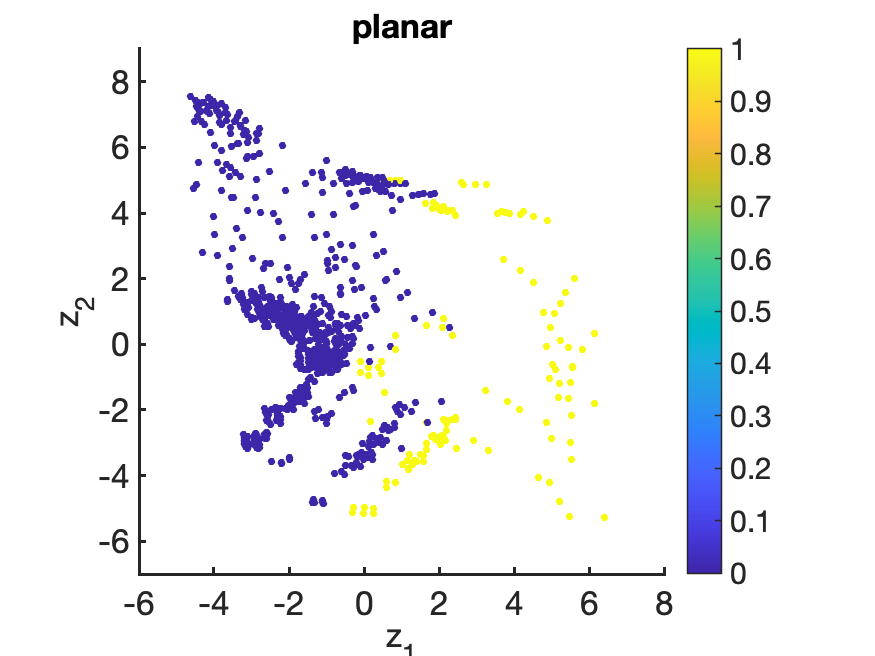}

\includegraphics[width=.22\linewidth]{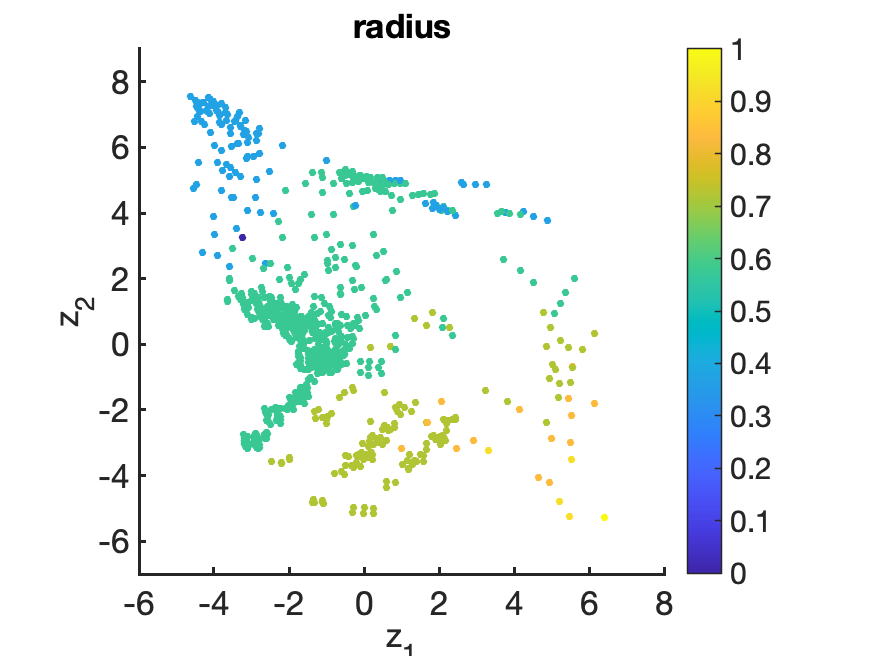}
\includegraphics[width=.22\linewidth]{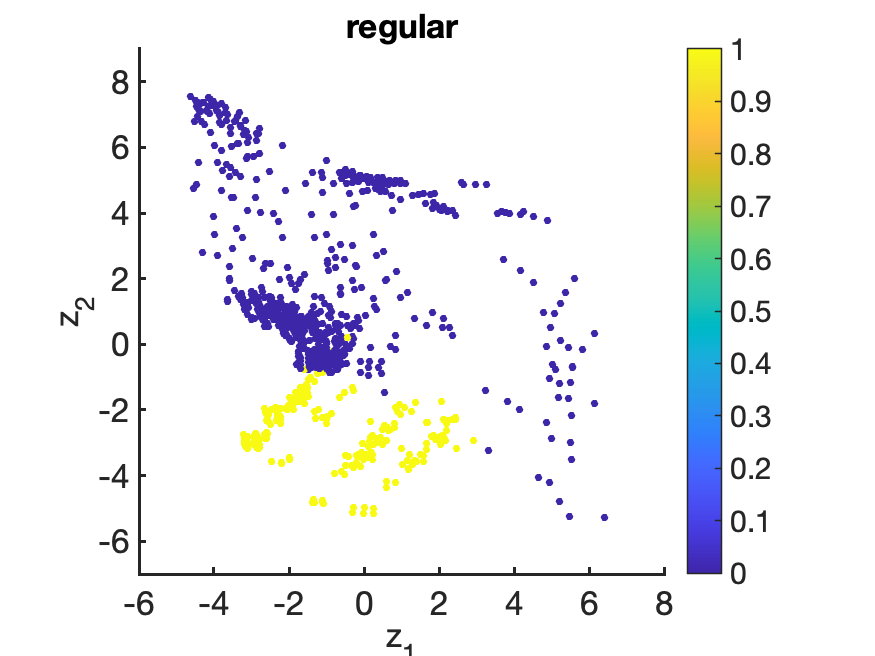}
\caption{Distribution of the 10 features selected by MATILDA across the Instance Space}
\label{fig:isa-feature-plots}
\end{figure}

Figure \ref{fig:isa-feature-plots} enables us to visualise the features selected in Equation (\ref{eq:projection-transformation}). We notice that the 10 features selected are fairly diverse and range from spectral features, structural features and also some of the symmetry-based features studied by \cite{Herrman2021}. Each feature plot has different density and distribution patterns, which can help us understand the relationship between instance classes and their properties. Firstly, for \textit{Average Distance}, the average length of the shortest paths between all pairs of vertices,  we observe a wide distribution across the entire instance space. Specifically, the ``Power Law Trees" graphs have a higher value for distance, which makes sense given their Power Law structure. \textit{Group Size} (a feature studied by \cite{Herrman2021}) appears to form a distinct cluster of low-values for ``Uniform Random" graphs. This is because the \textit{Group Size} is the size of the automorphism group of $G$; as ``Uniform Random" graphs have edges connected based on a fixed probability $p$, the random structure means it is unlikely that the graph will have a symmetrical structure, leading to a lower value for \textit{Group S1ize}. Because ``Geometric graphs" are created by placing vertices in a Euclidean plane and connecting the pairs of vertices based on a radius measure, these graphs can exhibit symmetrical patterns due to their construction. ``Nearly Complete BiPartite graphs" also have an obvious symmetry because their vertices are interchangeable while maintaining graph adjacency structure. The following features exhibit similar distributions: the \textit{largest eigenvalue of the Laplacian}, often referred to as the \textit{spectral radius},  \textit{minimum degree} and \textit{number of edges}. Both ``4-Regular" and ``3-Regular" graphs exhibit a consistent distribution, which is explained by the uniform degree across nodes. ``Power-Law Tree" graphs have clustered to small values, which is consistent as trees do not have cycles and hence lower connectivity -- this would produce graphs with the property of low values for all three features. The ``Nearly Complete BiPartite" and ``Watts-Strogatz Small World"  graphs both have higher values for the spectral radius, which is reflective of their high connectivity. However, for \textit{minimum degree} and \textit{number of edges}, they have lower values. It is noteworthy that both ``Geometric" and ``Uniform Random" graphs demonstrate a high degree of variability across each of these three features, which is expected due to the stochastic nature of their generation process. The \textit{minimum dominating set}, a feature that is a measure of coverage in terms of the vertex influence, has a distinct distribution across the instance space. We notice in the left half of the instance space there are lower values, where ``3-regular", ``4-regular", ``Watts-Strogatz Small World", ``Uniform Random" and many ``Geometric" graphs exist.  The \textit{planarity} of a graph indicates whether or not a graph can be drawn on a plane without any edge crossing. Its distribution closely follows that of the minimum degree. The \textit{radius} feature, a measure of the minimum \textit{eccentricity} of any vertex in the graph, shows that there is a diverse range of values in the instance space. We can observe that graphs with smaller values are those with a more centralised structure (such as ``Watts-Strogatz Small-World" graphs and ``Geometric" graphs). Graphs with a high radius value are ``Power Law Tree" graphs, which is reflective of their structure. The final selected feature, $regular$, provides clear visual evidence that most instances actually exist in totally different parts of the instance space. This plot confirms that existing research has focussed on a very narrow set of instances and has not captured the diversity of instances adequately. In the next section, we explore the relationship between the performance of various initialisation strategies across the Instance Space. 

\subsection{Performance of QAOA Algorithms (Initialisation Strategies)}
\begin{figure}[h!]
\centering
\includegraphics[width=.45\linewidth]{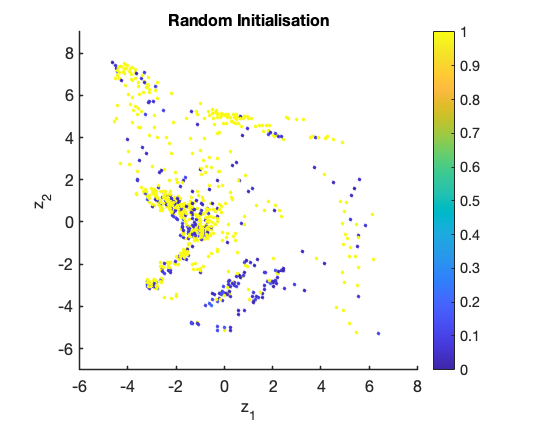}
\hfill
\includegraphics[width=.45\linewidth]{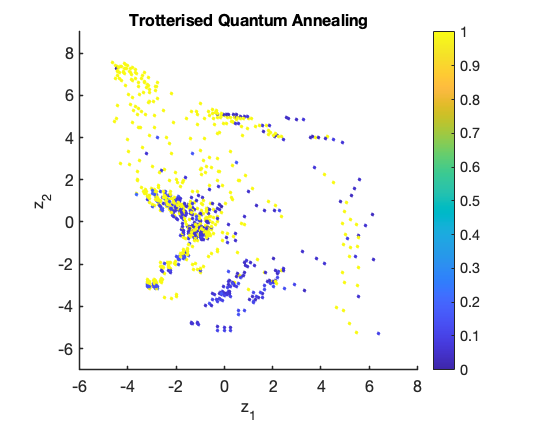}
\includegraphics[width=.45\linewidth]{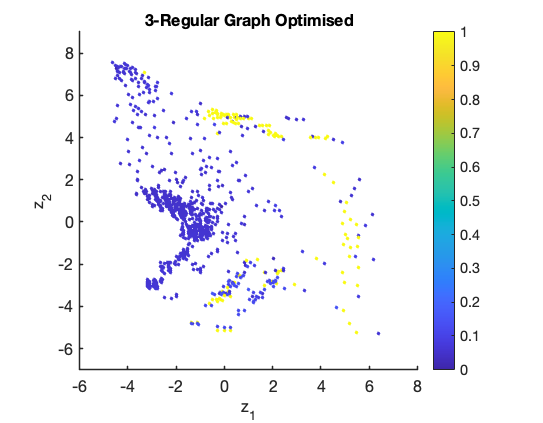}
\hfill
\includegraphics[width=.45\linewidth]{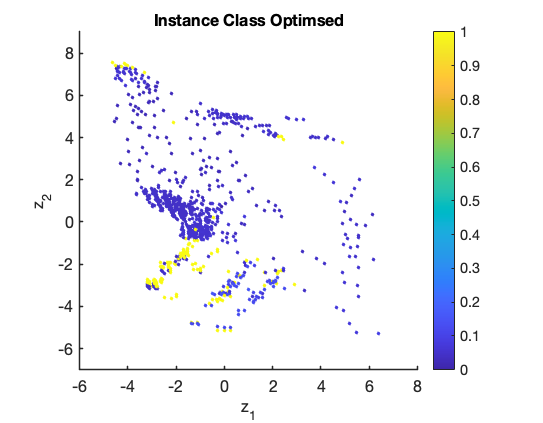}
\caption{Comparison of actual performance (normalised individually) for each algorithm: random initialisation (top-left), TQA (top-right), 3-regular graph optimised (bottom-left), and instance class optimised (bottom-right).}
\label{fig:isa-perf-plots-individually-normalised}
\end{figure}

We will now present the Instance Space with a view of the performance of initialisation algorithms. We have defined our performance measure in Section \ref{sec:isa-performance-space} as the number of function evaluations taken to achieve an acceptable approximation ratio $\alpha$. Figure \ref{fig:isa-perf-plots-individually-normalised} visualises the performance of each initialisation algorithm across the instance space. The scale has been normalised for each individual algorithm (to see the same graph, normalised globally, see Figure \ref{appendix:global-alg-perf}). A value closer to 1 indicates poor performance, and a value closer to 0 indicates better performance. For ``Random Initialisation", we observe that the performance is highly variable, with many instances showing poor performance. However, it does seem to perform better on ``3-Regular Graph" instances when compared to the rest of the instance space. \textit{Trotterised Quantum Annealing}, shows a concentration of better performance for ``3-Regular Graphs". The algorithm \textit{3-Regular Graph Optimised}, which uses $(\gamma, \beta)$ parameters that are pre-optimised for 3-regular graphs as initial values for all instance classes, performs surprisingly well across the instance space -- especially in the regions of the instance space that are occupied by ``3-Regular Graphs", ``4-Regular Graphs", ``Uniform Random", ``Watts Strogatz Small World" graphs and ``Geometric" graphs. This finding does provide some evidence to the claim that parameters optimised for one instance class can be transferred to another; this result is consistent with the findings in \cite{Galda2021}. However, we see a distinct performance decline in regions occupied by ``Nearly Complete BiPartite" graphs and ``Power Law Tree" graphs. Based on Figure \ref{fig:isa-feature-plots}, we can see that both these classes share low values for minimum degree, are planar and are not regular in their structure. Finally, for the \textit{Instance Class Optimised} initialisation strategy, we observe robust performance across the instance space. It performs particularly well in the area region of the instance space where other strategies perform poorly. From Figure \ref{fig:source-dist} we can see this is where ``Power Law Tree" graphs and ``Nearly Complete BiPartite" graphs exist.

\begin{figure}[h!]
\centering
\includegraphics[width=.45\linewidth]{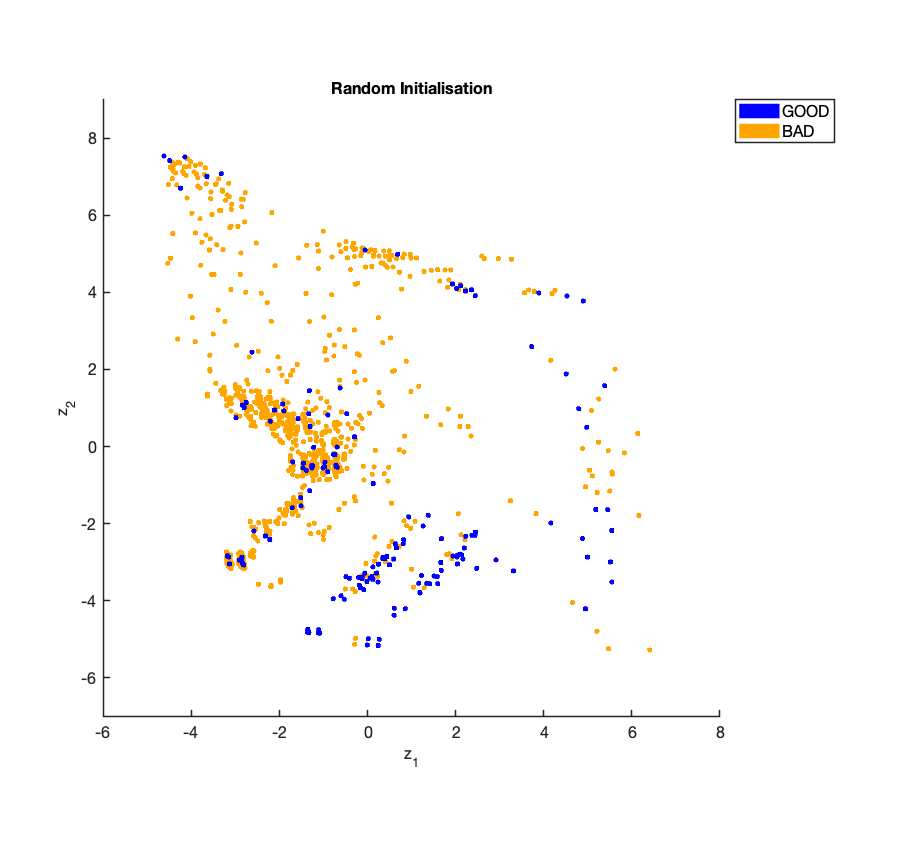}
\hfill
\includegraphics[width=.45\linewidth]{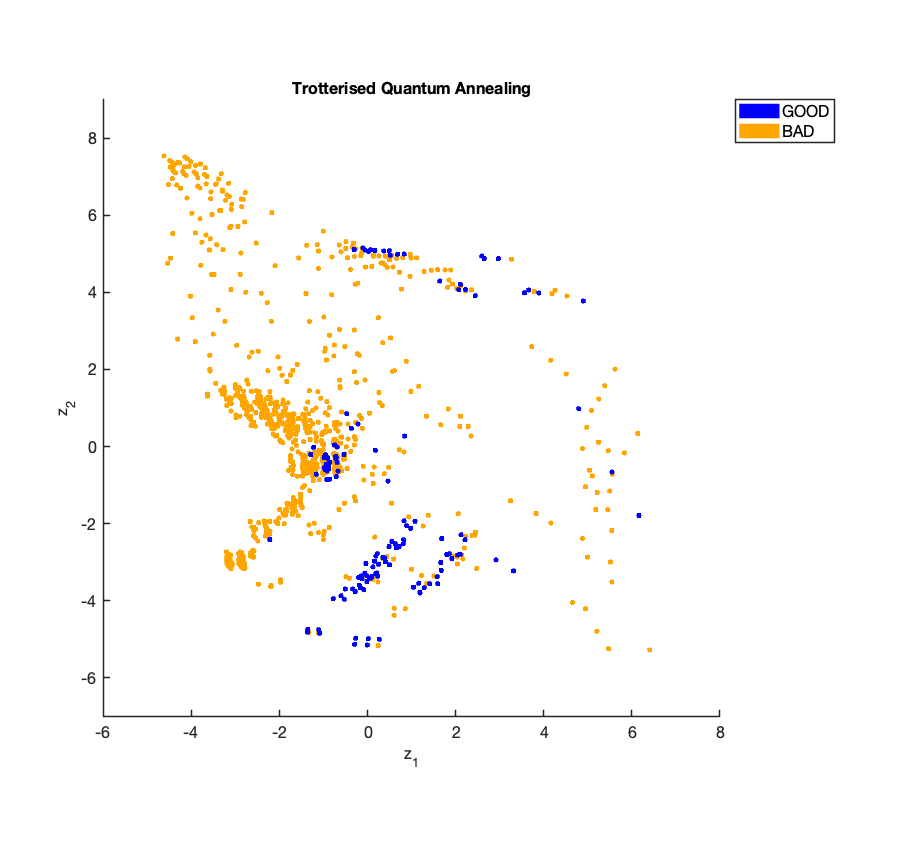}
\includegraphics[width=.45\linewidth]{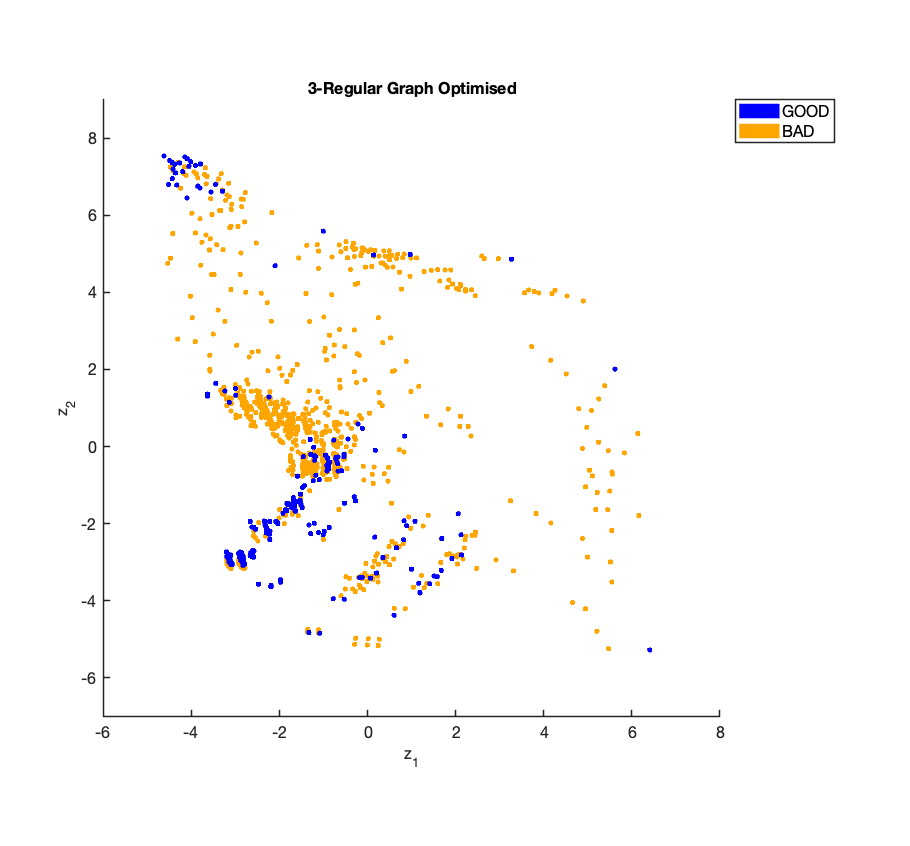}
\hfill
\includegraphics[width=.45\linewidth]{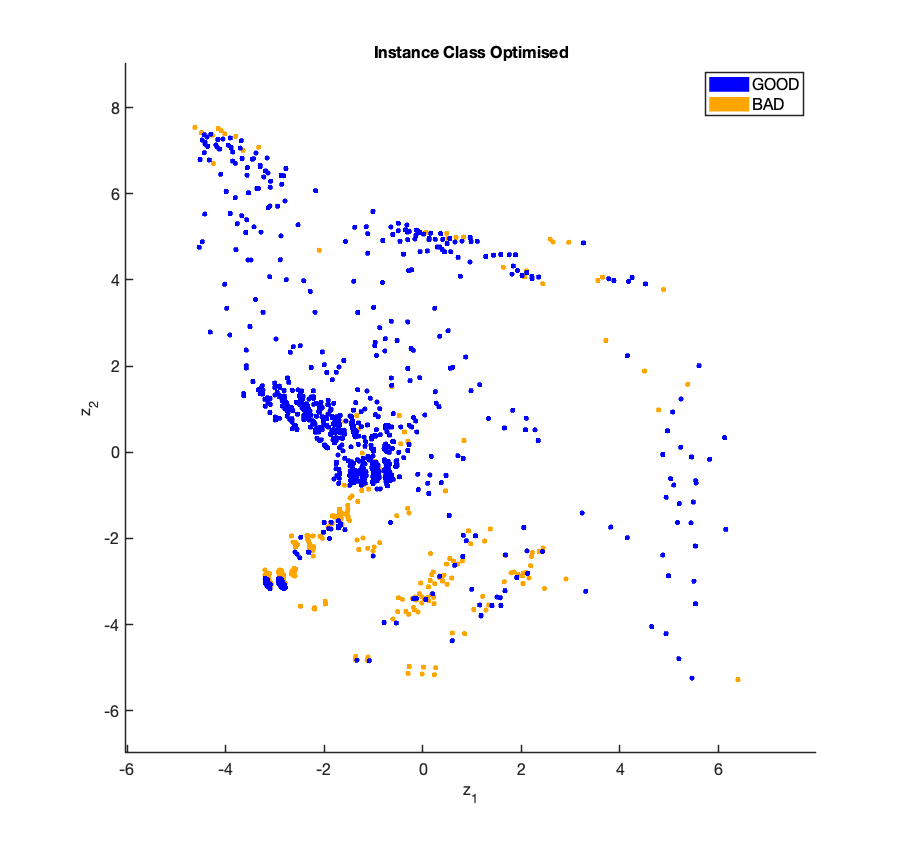}
\caption{Comparison of binary performance for each ``algorithm" (QAOA initialisation strategy): random initialisation (top-left), TQA (top-right), 3-regular graph optimised (bottom-left), and instance class optimised (bottom-right).}
\label{fig:isa-binary-perf-plots}
\end{figure}

Figure \ref{fig:isa-binary-perf-plots} compares the binary performance of each initialisation algorithm. As discussed in the previous section, we define ``good" performance when an algorithm is within 10\% of the best-performing algorithm. In Figure \ref{fig:isa-binary-perf-plots} we observe that for Random Initialisation, the distribution of ``good" outcomes is dispersed across the instance space, suggesting a lack of consistent convergence towards optimal solutions. This strategy, while simple, appears to have significant variability in performance, which may be due to the stochastic nature of the initial parameter choices. Similarly, the TQA strategy, depicted in the top-right quadrant, shows a similar distribution of ``good" outcomes. However, as we later see in Figure \ref{fig:footprint-prediction-plots}, the relative performance of the TQA algorithm for ``3-Regular Graphs" is superior. Based on Figure \ref{fig:isa-feature-plots} we can also see instances that are `regular' and `planar' exist in the regions where TQA performs well. The \textit{3-Regular Graph optimised} parameter initialisation method, located in the bottom-left quadrant, demonstrates varying levels of performance. It optimises regular graphs effectively but does not perform well in other areas of the instance space. The Instance Class Optimised method, shown in the bottom-right quadrant, demonstrates a ``good" performance, largely across the entire instance space, except we note that for ``4-Regular" graphs, the instance class optimised parameters work less well than simply using those of ``3-Regular" graphs, this is likely due to the fact the $(\gamma, \beta)$ parameters we used were experimentally computed for 8-node graphs and then seeded as initial parameters for a 12-node graph. Otherwise, we note that the instance class optimised method performs by far the best. We hypothesize that this is due to the tailored approach of the instance class initialisation and the robustness of those parameters when increasing graph size.   Figure \ref{fig:footprint-prediction-plots} highlights the substantial impact that instance characteristics can have on the success of quantum algorithms. These results provide evidence that the parameter optimisation landscape does, in fact depend on the instance class and characteristics of instances. It also shows that the effectiveness of parameter initialisation strategies depends on instance characteristics. However, it is important also to note that the parameters $(\gamma, \beta)$ used for this QAOA were optimised for 8-node graphs (rather than 12-node graphs). Along with the parameter concentration observed in Figure \ref{fig:histogram-of-optimal-parameters-by-instance-class}, these results suggest that within an instance class, we can transfer parameters from smaller instances to large instances (in our case from $n=8$ to $n=12$). This also means that an effective strategy could be to optimise $(\gamma, \beta)$ on smaller instances and then utilise those same parameters in large instances. This result was also shown by \cite{Galda2021} but was only applied to random graphs and $d-$regular graphs. However, our results also provide evidence that the optimal choices of $(\gamma, \beta)$ are not instance-independent \textit{between} instance classes. We can see this clearly in Figure \ref{fig:algo-perf-on-single-instance} and Figure \ref{appendix:global-alg-perf} where using ``3-regular" graph parameters on ``Power Law" graph instances and ``Nearly Complete Bipartite" graph instances yields in poor performance.

\subsection{Algorithm Footprints}
\begin{figure}[h!]
    \centering
    \begin{subfigure}{0.45\textwidth}
        \includegraphics[width=\textwidth]{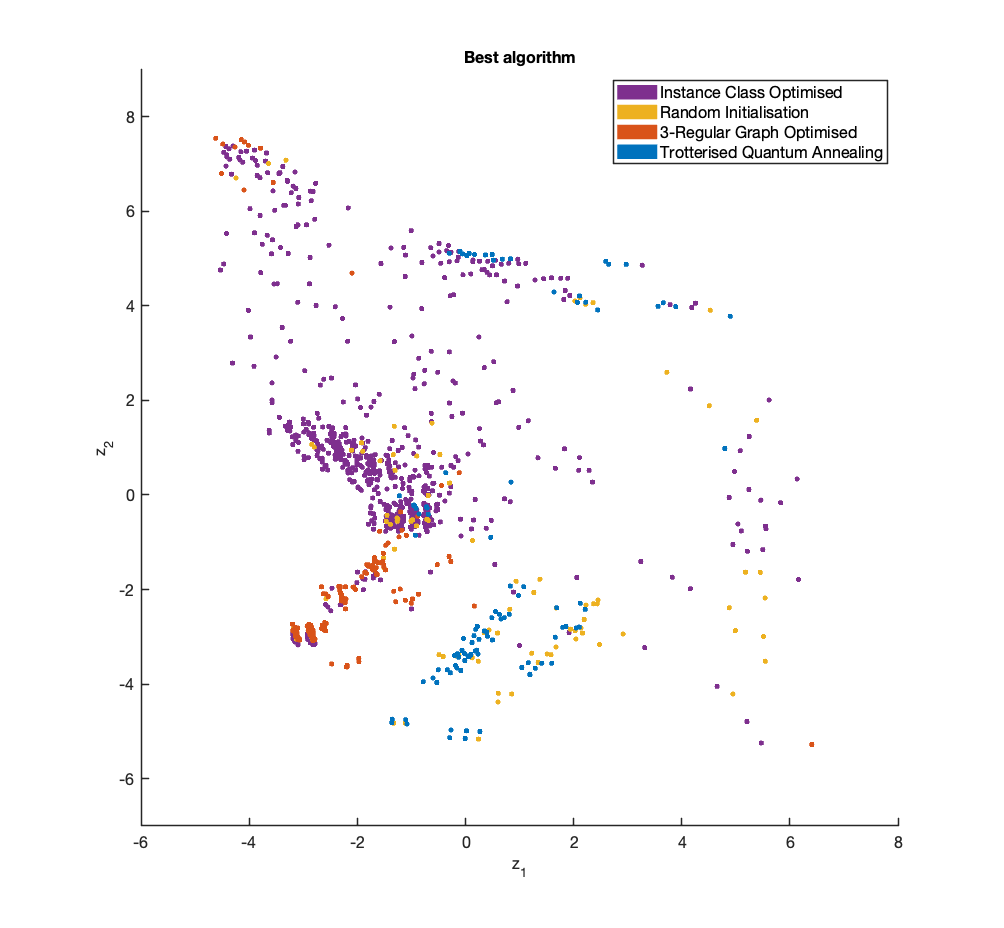}
        \label{fig:best-algorithm}
    \end{subfigure}
    \hfill
    \begin{subfigure}{0.45\textwidth}
        \includegraphics[width=\textwidth]{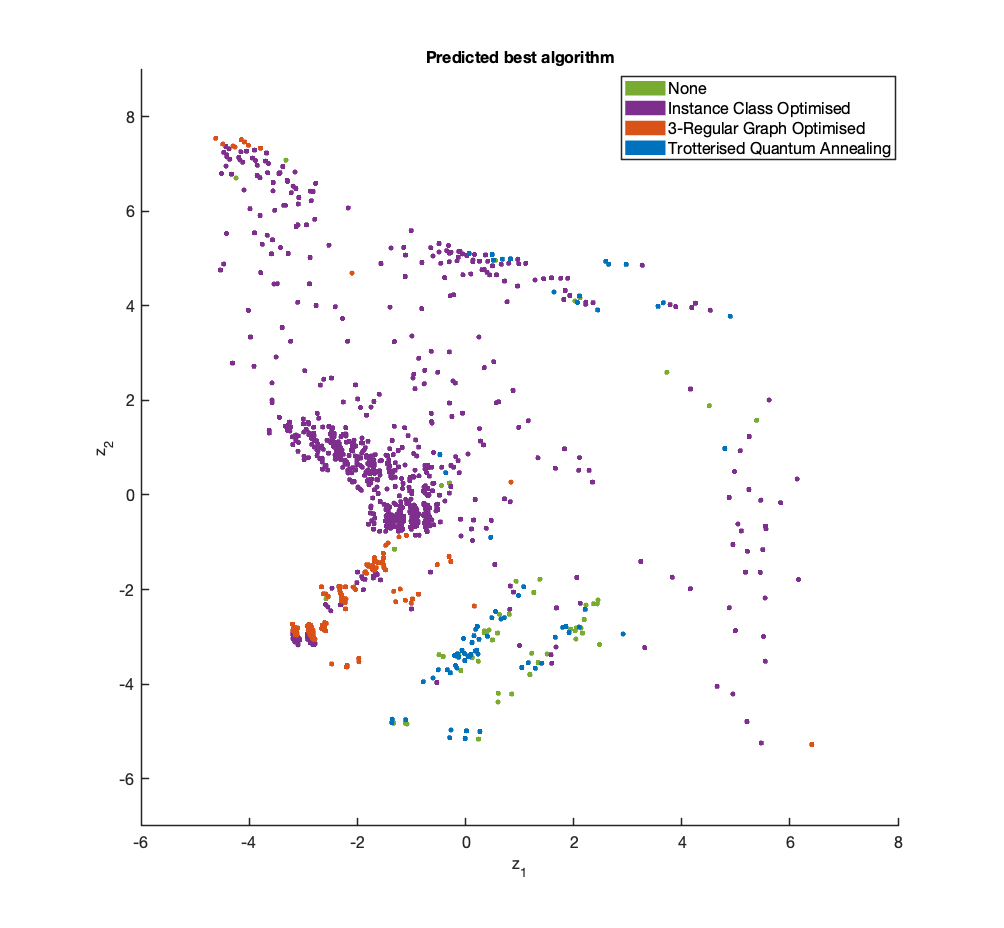}
        \label{fig:predicted-best-algorithm}
    \end{subfigure}
    \hfill
    \begin{subfigure}{0.45\textwidth}
        \includegraphics[width=\textwidth]{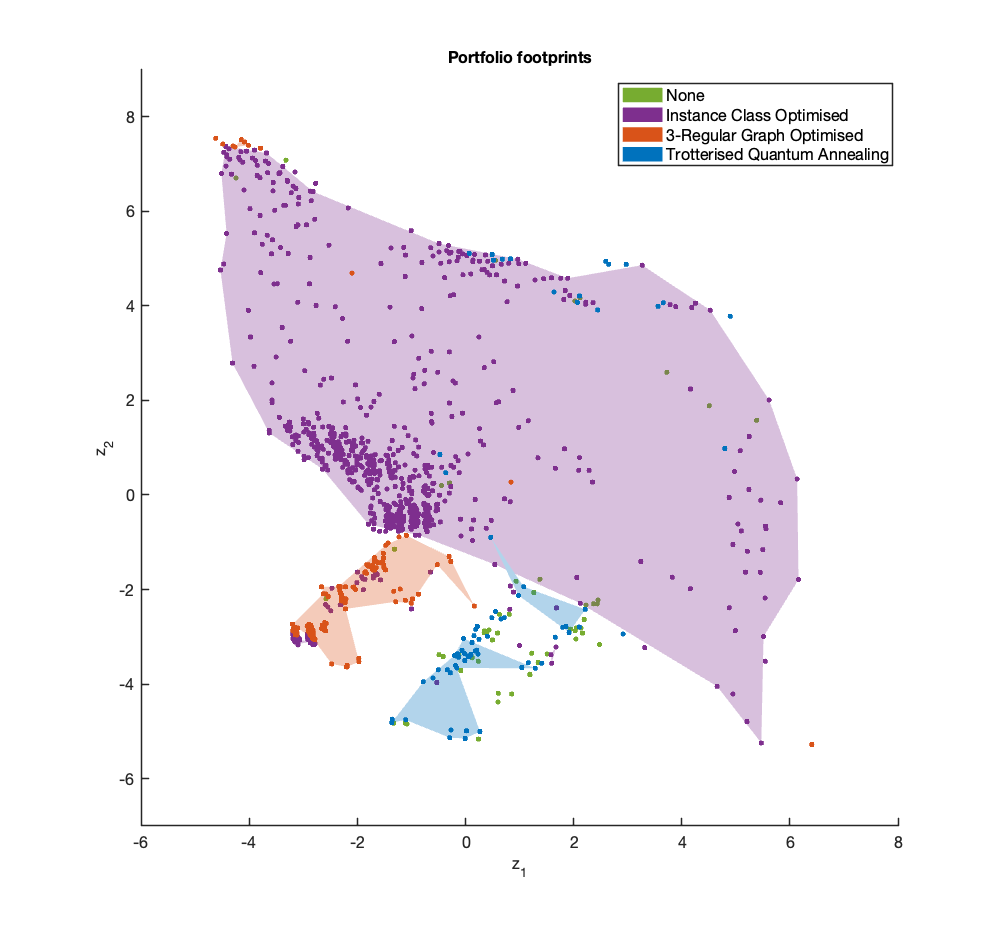}
        \label{fig:footprint-portfolio}
    \end{subfigure}
    \caption{Comparison of best algorithm (top-left), predicted best algorithm (top-right) and algorithm footprints (bottom-left). }
\label{fig:footprint-prediction-plots}
\end{figure}

\begin{figure}[h!]
\centering
\includegraphics[width=.45\linewidth]{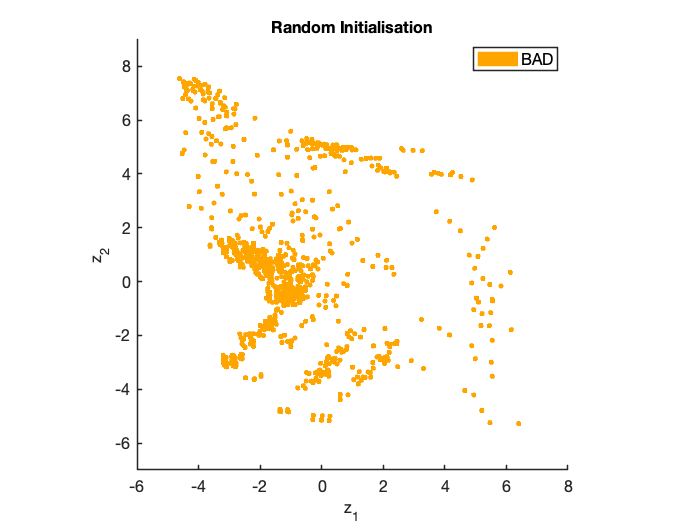}
\hfill
\includegraphics[width=.45\linewidth]{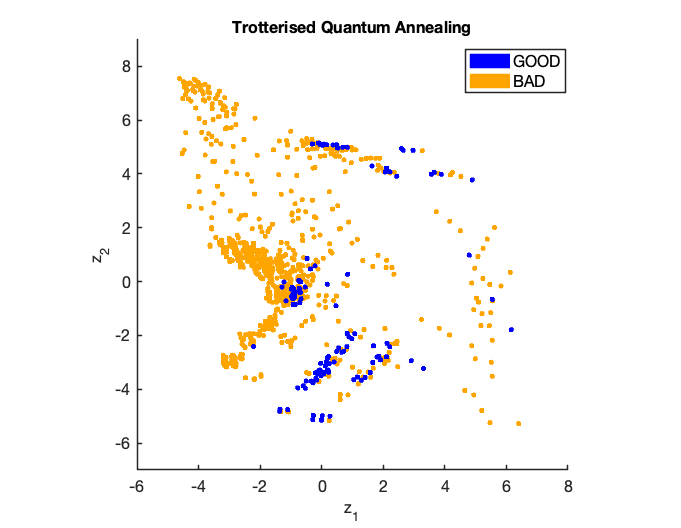}
\includegraphics[width=.45\linewidth]{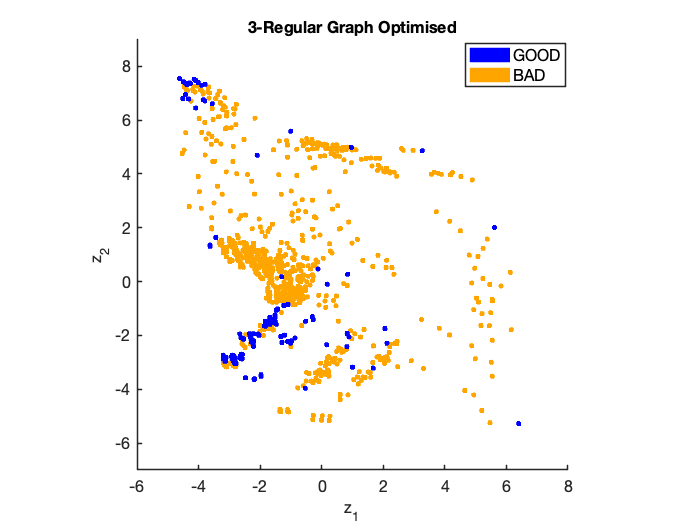}
\hfill
\includegraphics[width=.45\linewidth]{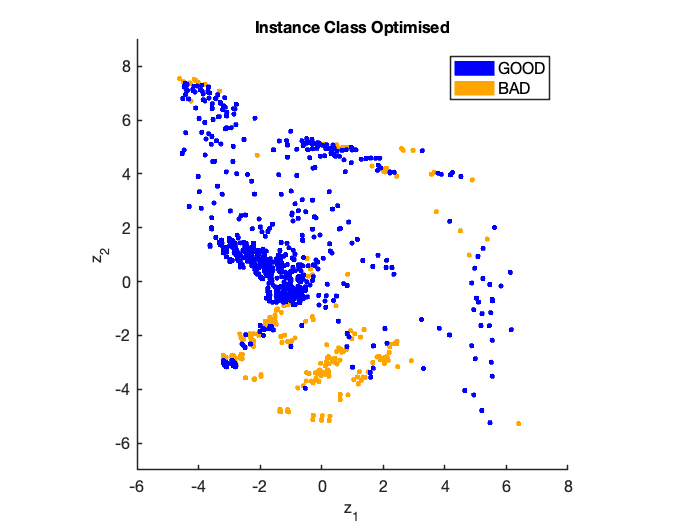}
\caption{Comparison of SVM Predictions of `GOOD` or `BAD` performance for each algorithm: random initialisation (top-left), TQA (top-right), 3-regular graph optimised (bottom-left), and instance class optimised (bottom-right).}
\label{fig:isa-svm-predictions}
\end{figure}

We can use binary classifications to analyze and visualise different regions of the instance space where each initialisation strategy is expected to perform well. To achieve this, we make use of the binary labels ``good" and ``bad". As explained in Section \ref{sec:isa-performance-space}, MATILDA employs an SVM with a gaussian kernel to learn the footprint of each algorithm. In Figure \ref{fig:isa-svm-predictions}, we visualise how the SVM predicts each algorithm would perform, and Figure \ref{fig:footprint-prediction-plots} compares the footprint region of each initialisation algorithm. The results presented in \ref{fig:footprint-prediction-plots} compare the actual best algorithm, the predicted best algorithm, and the algorithm footprint defined by MATILDA. The figure clearly shows that the algorithms optimized for ``Instance Class" hold the largest footprint across $z_1$ and $z_2$. Surprisingly, the ``3-regular graph optimized" initialisation strategy footprint occupies the region where 4-regular graphs exist, and for 3-regular graphs, TQA has been selected. We can see in Figure \ref{fig:footprint-prediction-plots} that the MATILDA SVM algorithm accurately captured the distribution of the best algorithm in comparison to its predicted version and algorithmic footprints.

\section{Conclusion}
In this paper, we have conducted the first instance space analysis of QAOA with a specific focus on parameter initialisation strategies for solving MaxCut problems. By exploring a more diverse set of instance classes our analysis has revealed that there is a relationship between instance characteristics and the efficacy of a given initialisation strategy. We have provided evidence that the parameter optimisation landscape depends on the instance class, and the effectiveness of different parameter initialisation strategies is also, therefore, dependent on instance characteristics. We have proposed an instance-based initialisation strategy that is more effective at finding optimal parameters, and achieving superior performance of QAOA on MaxCut, compared to alternatives that initialise the search undertaken by the ADAM algorithm in an instance independent manner. The resulting optimised QAOA is able to achieve an acceptable approximation ratio in a reduced number of iterations when initialised appropriately.

Using a comprehensive set of 2,100 graphs across seven instance classes has afforded us the opportunity to note the diversity of the constructed instance space, and how limited our conclusions would be if we only considered the 3-regular, 4-regular and random graphs so prevalent in the literature. Nevertheless, despite considerably augmenting the instance space, we still observe significant gaps in the instance space,  particularly around the regions occupied by ``Power Law Tree" graphs and ``Nearly Complete Bipartite" graphs. In our future  research we will aim to generate a wider variety of instances within these regions of the instance space. Techniques to evolve instances are discussed in \cite{Munoz2020}. The current analysis also focussed on only four initialisation strategies, but as discussed in Section \ref{sec:methods}, there have been many approaches to parameter initialisation techniques introduced in recent literature. Of course, there are also many alternatives to ADAM as an optimisation algorithm for the parameter search, and ISA could be applied to address the question of how the combination of algorithm, initialisation and parameters interplay with instance characteristics to impact QAOA performance. A more comprehensive comparison is left to future work, extending on the foundations of this initial case study.

Other areas worthy of further investigation are performing ISA on actual IBM-Q hardware and understanding how error distributions on real quantum machines are influenced by instance characteristics. Expanding this work to study quantum algorithm performance on other optimisation problems beyond $\text{MaxCut}$ is another possible extension with some suitable benchmark problems identified in recent work \citep{abbas2023quantum}. Lastly, extending this analysis to test other VQEs is another possible avenue for exploration. The library problems contained on the MATILDA website \citep{Smith-Miles2020} include many useful features for other problems to support such analysis.

In conclusion, by applying instance space analysis for the first time to QAOA implementations for solving  $\text{MaxCut}$, we have not only gained valuable insights into the question of whether instance characteristics impact the optimal QAOA parameters. We have introduced a new technique to initialise QAOA parameters, QIBPI. Lastly and most importantly, we have laid the foundations for demonstrating how ISA can be used to address many other open questions in the field of quantum computing, where it is currently unclear how instance properties impact quantum algorithm performance.

\ACKNOWLEDGMENT{%
 This research was supported by the Australian Research Council under grant number IC200100009 for the ARC Training Centre in Optimisation Technologies, Integrated Methodologies and Applications (OPTIMA). The first author is also supported by a Research Training Program Scholarship from The University of Melbourne. The authors gratefully acknowledge the IT infrastructure support provided by The University of Melbourne’s Research Computing Services and the Petascale Campus Initiative. 
}

%
\begin{APPENDIX}{}
\section{Instance Definitions}\label{appendix:defs} 
\setcounter{definition}{0}


\begin{definition}[Uniform Random Graph]
A Uniform Random Graph, denoted as $G_{n,p}$, is defined for a set of $n$ vertices such that every pair of distinct vertices is connected by an edge with probability $p$ independently of every other edge. Formally, for every pair of vertices $v_i, v_j \in G_{n,p}$, where $i \neq j$, we have:
\[ P((v_i, v_j) \in E(G_{n,p})) = p \]
where $E(G_{n,p})$ denotes the edge set of $G_{n,p}$.
\end{definition}

\begin{definition}[Power Law Tree]
A Power Law Tree is a graph that is both a tree and follows a power law degree distribution. For a given degree $k$, the probability $P(k)$ that a vertex in the tree has degree $k$ is proportional to $k^{-\gamma}$ for some constant $\gamma > 1$. This implies a small number of vertices with very high degree (hubs) and a large number of vertices with low degree.
\end{definition}

\begin{definition}[Watts-Strogatz Small World Graph]
A Watts-Strogatz Small World Graph, denoted as $G_{n,k,\beta}$, is a regular graph initially constructed by arranging $n$ vertices in a ring, each connected to $k$ nearest neighbours, and then randomly re-wiring each edge with probability $\beta$ to a new vertex, such that self-loops and duplicate edges are avoided. This model interpolates between a regular lattice ($\beta = 0$) and a random graph ($\beta = 1$), capturing the small-world phenomenon of most real networks.
\end{definition}

\begin{definition}[Geometric Graph]
A Geometric Graph, denoted as $G_{n,r}$, is defined for a set of $n$ vertices, where each vertex corresponds to a random point in a metric space, and edges are formed between vertices that are within a distance $r$ from each other. Formally, for every pair of vertices $v_i, v_j \in G_{n,r}$:
\[ (v_i, v_j) \in E(G_{n,r}) \Leftrightarrow d(v_i, v_j) \leq r \]
where $d(\cdot, \cdot)$ is a distance function in the given metric space. Figure \ref{fig:geometric-algorithm} provides an illustrative example
\end{definition}

\begin{figure}[htp]
    \centering
    \includegraphics[width=0.9\linewidth]{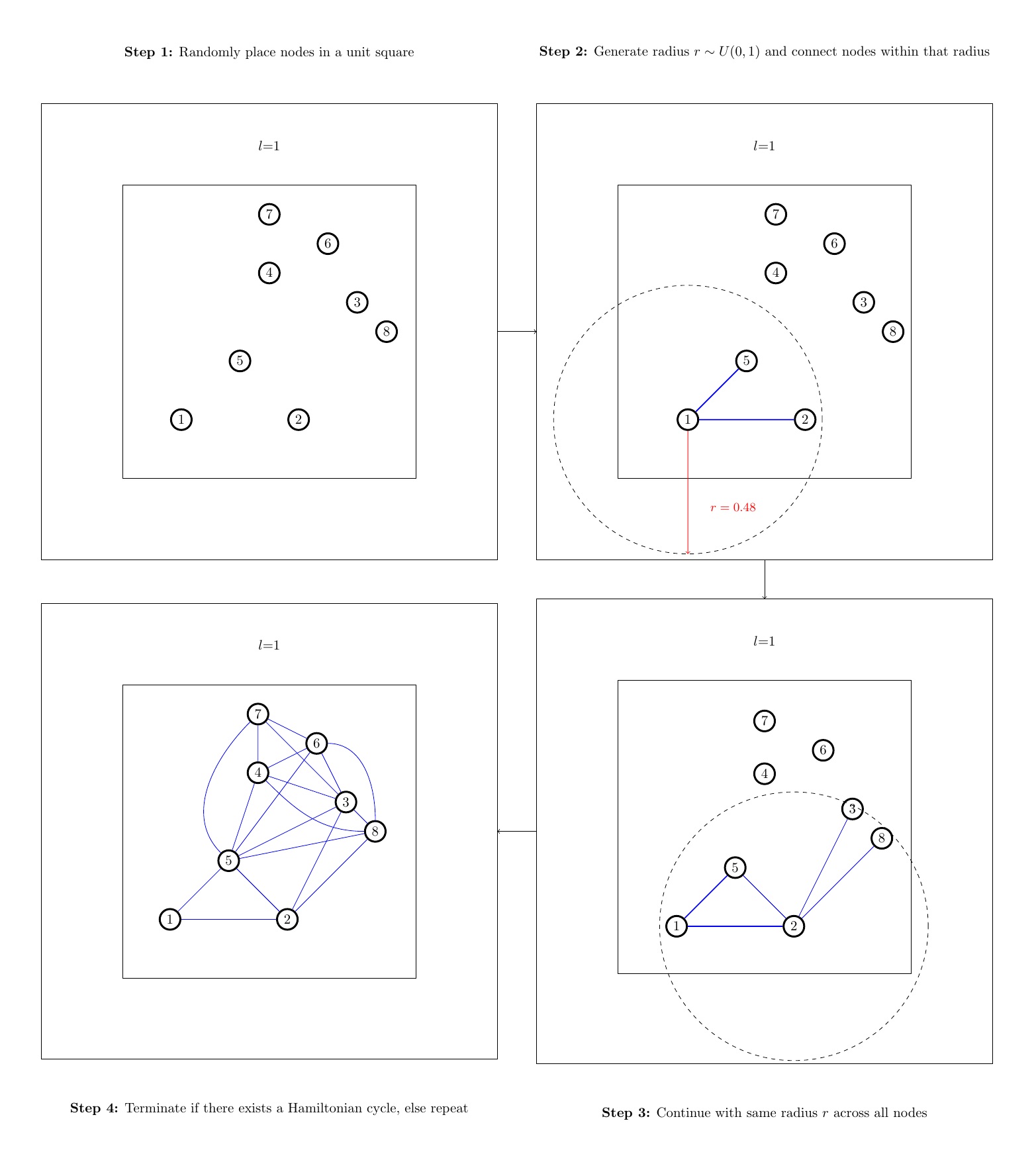}
    \caption{Example of a geometric graph $G$ with 8 nodes}
    \label{fig:geometric-algorithm}
\end{figure}

\begin{definition}[Regular Graph]
A $k$-Regular Graph, denoted as $G_{n,k}$, is a graph where each of the $n$ vertices has exactly $k$ neighbors. In other words, the degree of every vertex in $G_{n,k}$ is $k$.
\end{definition}

\begin{definition}[Nearly Complete Bipartite Graph]
A Nearly Complete Bipartite Graph refers to a bipartite graph $G=(V_1, V_2, E)$ in which $V_1$ and $V_2$ are the two disjoint vertex sets such that every vertex in $V_1$ is connected to almost every vertex in $V_2$ and vice versa, except for a few missing edges that prevent the graph from being a complete bipartite graph.
\end{definition}

\section{Feature Definitions}\label{appendix:feat-defs} 

In this section, we define the $\text{MaxCut}$ features that appear in this paper. These definitions are given a graph $G = (V, E)$.

\begin{definition}[Instance Size]
The instance size of a graph $G$, denoted as $|V|$, is the number of vertices in $G$. Also referred to as the number of vertices or nodes.
\end{definition}

\begin{definition}[Number of Edges]
The number of edges in a graph $G$, denoted as $|E|$, is the total count of edges connecting the vertices in $G$.
\end{definition}

\begin{definition}[Bipartite Graph]
A graph $G$ is bipartite if its vertex set can be divided into two disjoint sets such that every edge connects a vertex in one set to a vertex in the other set.
\end{definition}

\begin{definition}[Clique Number]
The clique number of a graph $G$, denoted as $\omega(G)$, is the number of vertices in the largest complete subgraph in $G$.
\end{definition}

\begin{definition}[Connected Graph]
A graph $G$ is connected if there is a path between every pair of vertices.
\end{definition}

\begin{definition}[Density]
The density of an undirected graph $G$, denoted as $\delta(G)$, is defined as the ratio of the number of edges $m$ in the graph to the maximum possible number of edges among the vertices. For a graph $G$ with $n$ vertices and $m$ edges, the density is given by: $\delta(G) = \frac{2m}{n(n-1)}$. The density is 0 for a graph without edges and 1 for a complete graph.
\end{definition}

\begin{definition}[Edge Connectivity]
The edge connectivity of a graph $G$, denoted as $\lambda(G)$, is the minimum number of edges that must be removed to disconnect $G$.
\end{definition}

\begin{definition}[Maximum Degree]
The maximum degree of a graph $G$, denoted as $\Delta(G)$, is the largest degree of any vertex in $G$.
\end{definition}

\begin{definition}[Minimum Degree]
The minimum degree of a graph $G$, denoted as $\delta(G)$, is the smallest degree of any vertex in $G$.
\end{definition}

\begin{definition}[Minimum Dominating Set]
A minimum dominating set for a graph $G$ is a smallest subset of vertices such that every vertex in $G$ is either in the subset or adjacent to a vertex in the subset.
\end{definition}

\begin{definition}[Regular Graph]
A graph $G$ is regular if all its vertices have the same degree.
\end{definition}

\begin{definition}[Smallest Eigenvalue]
The smallest eigenvalue of the adjacency matrix of a graph $G$.
\end{definition}

\begin{definition}[Vertex Connectivity]
The vertex connectivity of a graph $G$, denoted as $\kappa(G)$, is the minimum number of vertices that must be removed to disconnect $G$.
\end{definition}


\begin{definition}[Acyclic Graph]
A graph $G$ is acyclic if it contains no cycles.
\end{definition}

\begin{definition}[Average Distance]
The average distance in a graph $G$ is the average length of the shortest paths between all pairs of vertices.
\end{definition}

\begin{definition}[Diameter]
The diameter of a graph $G$, denoted as $\text{diam}(G)$, is the longest shortest path between any two vertices in $G$.
\end{definition}

\begin{definition}[Eulerian Graph]
A graph $G$ is Eulerian if it contains an Eulerian circuit, a circuit that visits every edge exactly once.
\end{definition}

\begin{definition}[Number of Components]
The number of components in a graph $G$, denoted as $\kappa'(G)$, is the number of connected subgraphs in $G$.
\end{definition}

\begin{definition}[Planar Graph]
A graph $G$ is planar if it can be drawn on a plane without any edges crossing.
\end{definition}

\begin{definition}[Eccentricity]
The eccentricity of a vertex $v$ is the maximum distance between $v$ and all other vertices.
\end{definition}

\begin{definition}[Radius]
The radius of a graph $G$, denoted as $\text{rad}(G)$, is the minimum eccentricity of any vertex in $G$.
\end{definition}


\begin{definition}[Algebraic Connectivity]
The algebraic connectivity of a graph $G$, denoted as $\lambda_2(L)$, is the second-smallest eigenvalue of its Laplacian matrix.
\end{definition}

\begin{definition}[Laplacian Matrix]
    The graph Laplacian matrix is the matrix $L = D-A$, where $A$ is the adjacency matrix and $D$ is the diagonal matrix of node degrees. The normalised Laplacian is given by $\hat{L} = D^{-1/2} L D^{1/2}$
\end{definition}

\begin{definition}[Laplacian Largest Eigenvalue]
The largest eigenvalue of the Laplacian matrix of a graph $G$, denoted as $\lambda_{\text{max}}(L)$.
\end{definition}

\begin{definition}[Laplacian Second Largest Eigenvalue]
The second largest eigenvalue of the Laplacian matrix of a graph $G$.
\end{definition}

\begin{definition}[Ratio of Two Largest Laplacian Eigenvalues]
The ratio of the two largest eigenvalues of the Laplacian matrix of a graph $G$.
\end{definition}

\begin{definition}[Ratio of Two Smallest Laplacian Eigenvalues]
The ratio of the two smallest eigenvalues of the Laplacian matrix of a graph $G$.
\end{definition}


\begin{definition}[Distance Regular]
A graph $G=(V,E)$ is distance regular if for any pair of vertices $u, v \in V$, the number of vertices that have a distance $i$ from $u$ is equal to the number of vertices that are distance $i$ from $v, \enspace \forall i \in \{1, \dots, d\}$. Where $d$ is the diameter of $G$
\end{definition}

\begin{definition}[Graph Automorphism]
A graph automorphism is a bijection $f: V \to V$ such that for any two vertices $u, v \in V$, $\{u, v\} \in E$ if and only if $\{f(u), f(v)\} \in E$.
\end{definition}

\begin{definition}[Automorphism Group]
The automorphism group of $G$, denoted $\text{Aut}(G)$, is defined as the set of all bijections $f: V \to V$ such that $\{u, v\} \in E$ if and only if $\{f(u), f(v)\} \in E$ for all $u, v \in V$.
\end{definition}

\begin{definition}[Group Size]
    The group size, or the order of $\text{Aut}(G)$, denoted $|\text{Aut}(G)|$, is the total number of such distinct automorphisms.
\end{definition}

\begin{definition}[Number of Cut Vertices]
   The number of cut vertices for a graph$G$. Where a cut vertex of a connected graph is a vertex whose removal disconnects the graph.
\end{definition}

\begin{definition} [Minimal Odd Cycle]
A \textit{minimal odd cycle} in a graph $G$ is an odd-length cycle that does not contain any other odd-length cycles. Formally, a cycle $C$ in $G$ is minimal odd if for every odd cycle $C'$ in $G$, $C' \not\subset C$.
\end{definition}

\begin{definition} [Orbit]
For a graph $G = (V,E)$. An orbit of a vertex $v \in V$ is the set of all vertices $a(v)$ where $a$ is $Aut(G)$.
\end{definition}

\section{Additional Instance Space Analysis Figures}

\begin{figure}[h!]
\centering
\includegraphics[width=.45\linewidth]{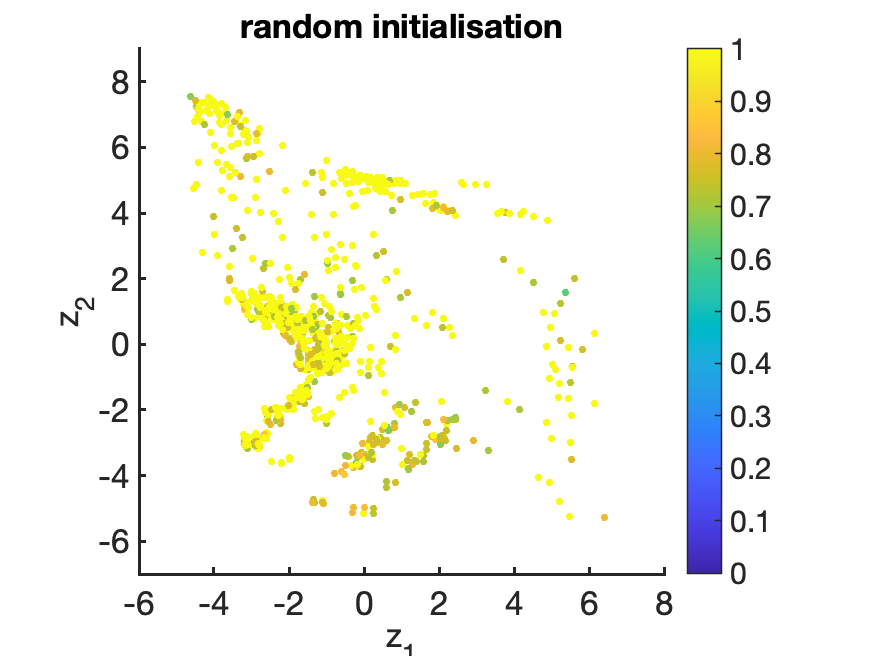}
\hfill
\includegraphics[width=.45\linewidth]{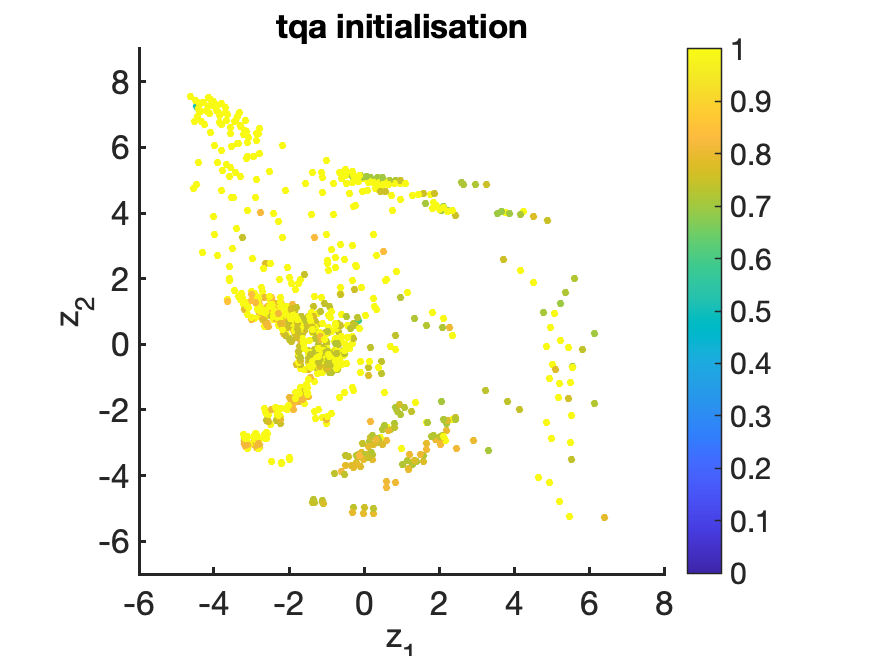}
\includegraphics[width=.45\linewidth]{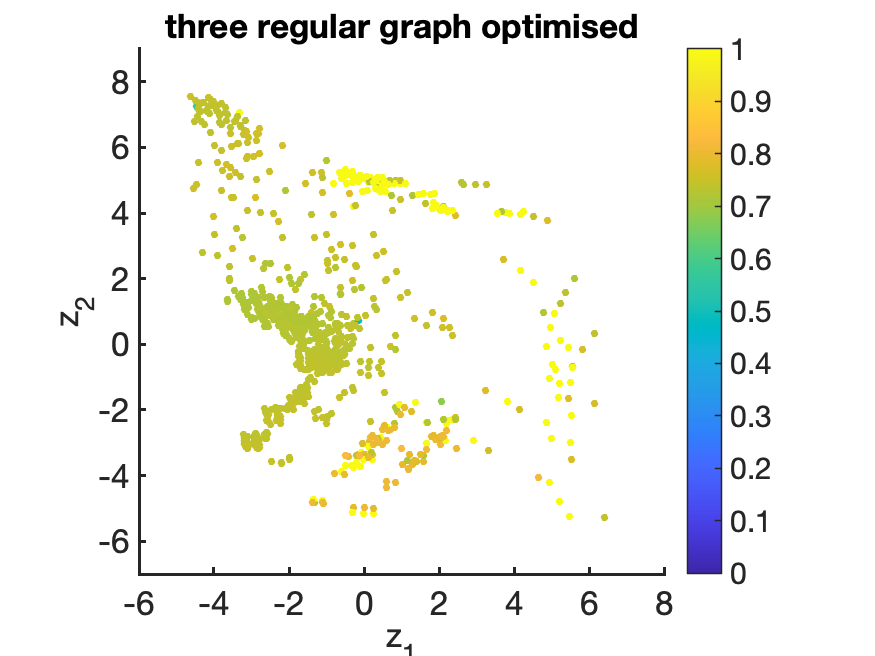}
\hfill
\includegraphics[width=.45\linewidth]{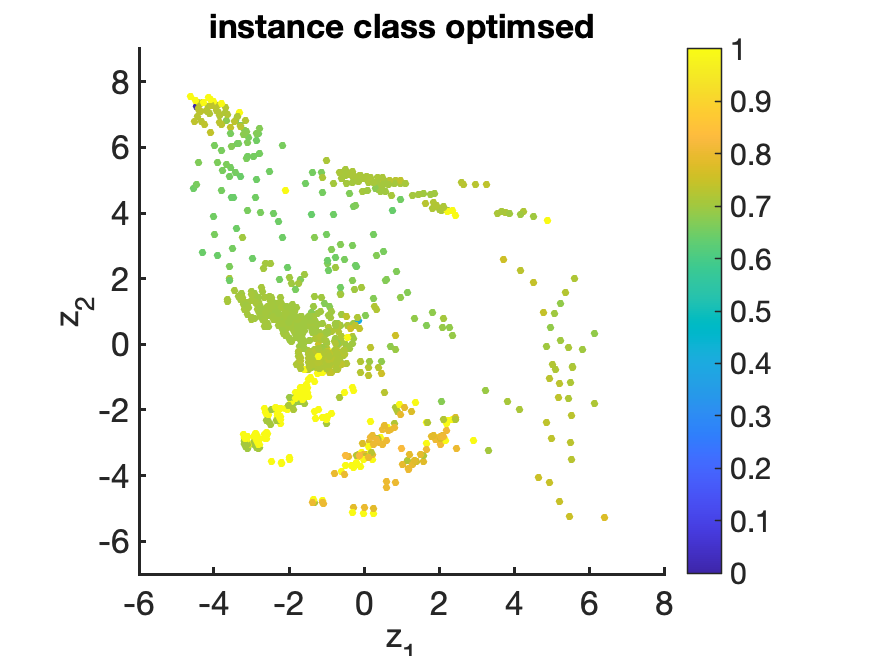}
\caption{Comparison of actual performance (normalised globally) for each algorithm: random initialisation (top-left), TQA (top-right), 3-regular graph optimised (bottom-left), and instance class optimised (bottom-right).}
\label{appendix:global-alg-perf}
\end{figure}

\begin{figure}[h!]
\centering
\includegraphics[width=.45\linewidth]{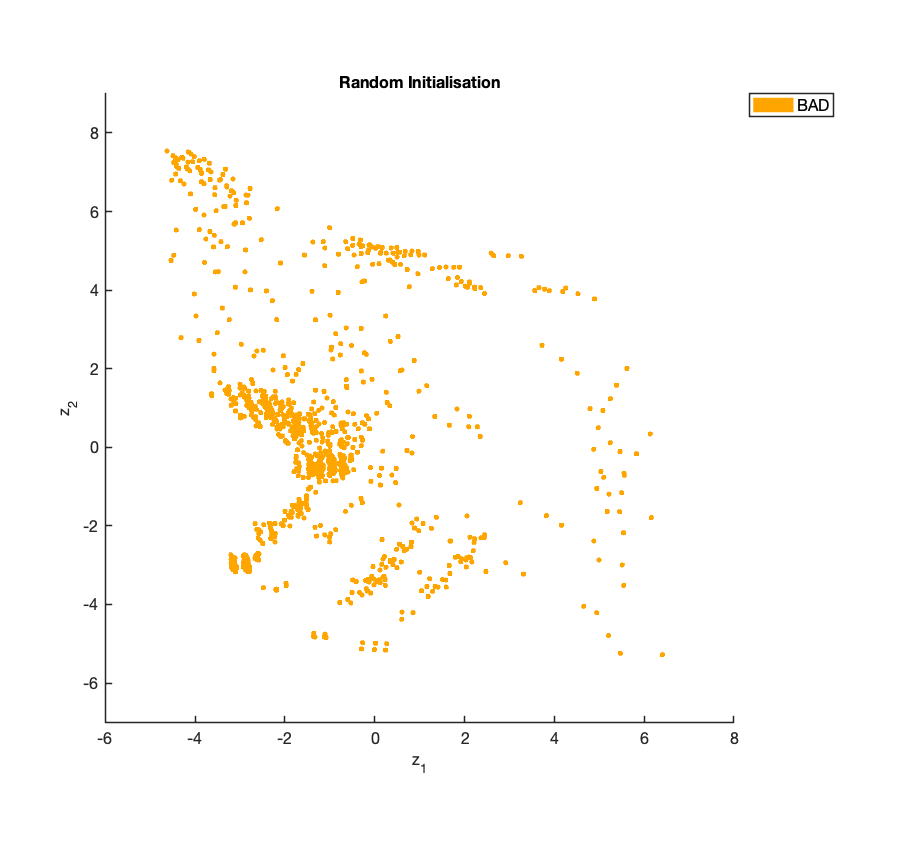}
\hfill
\includegraphics[width=.45\linewidth]{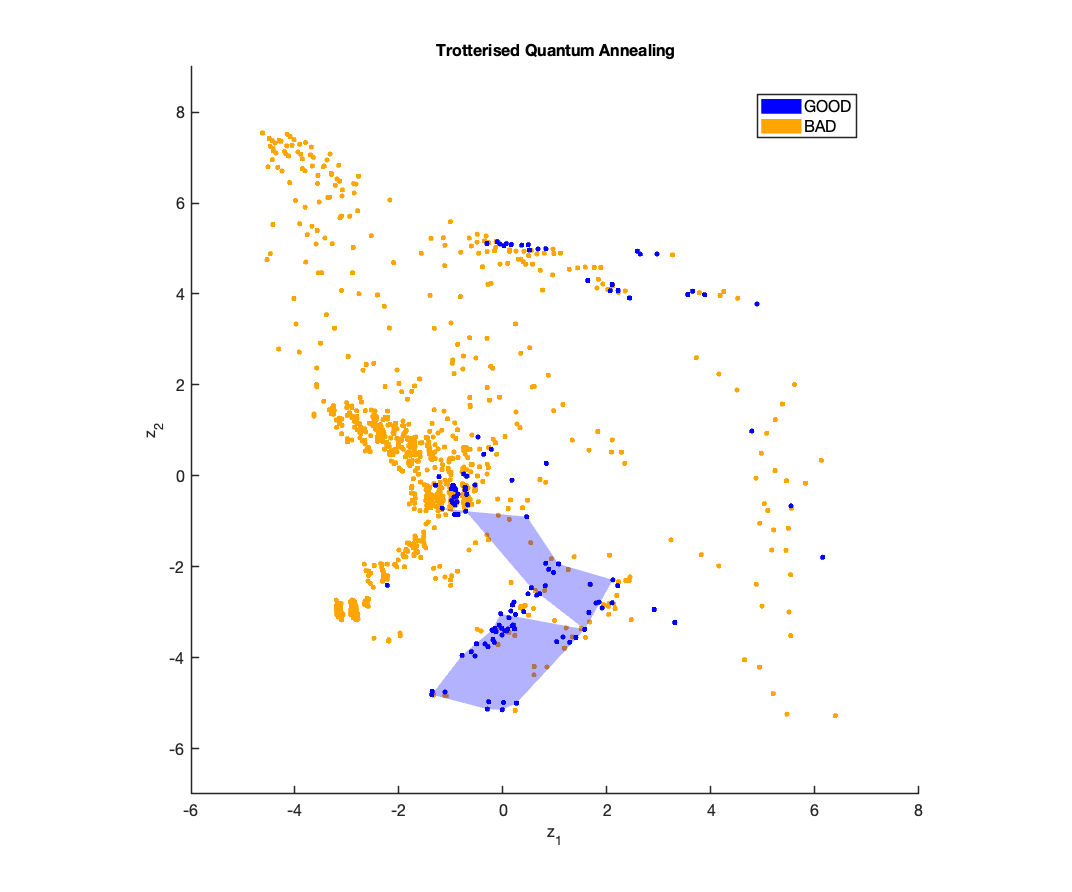}
\includegraphics[width=.45\linewidth]{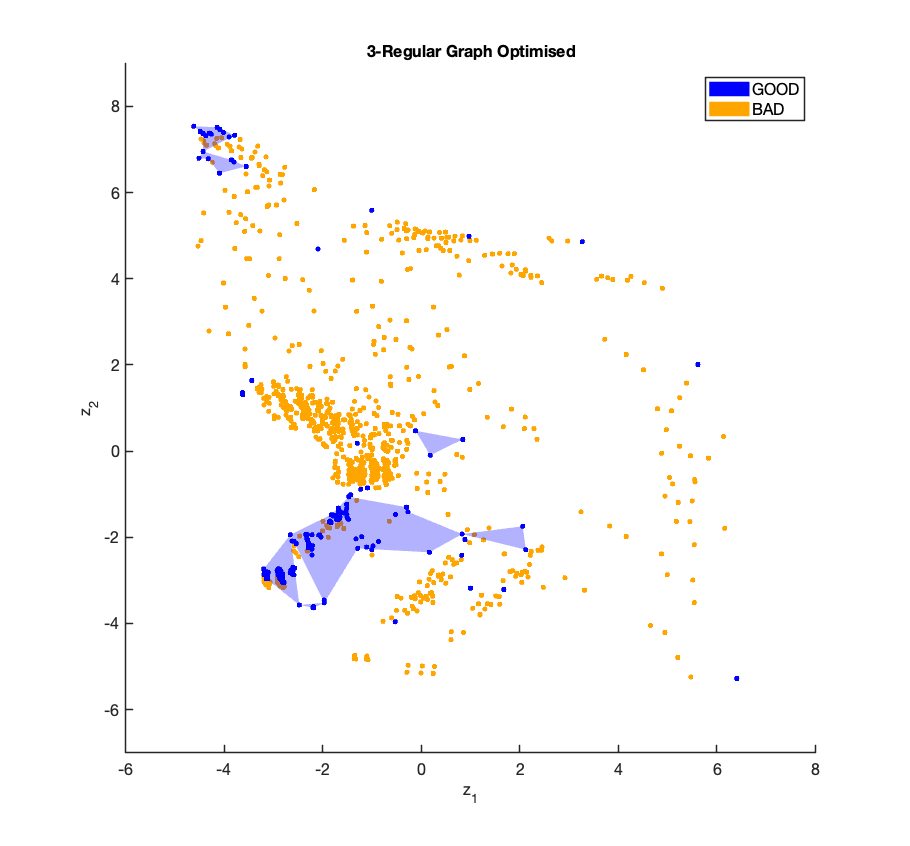}
\hfill
\includegraphics[width=.45\linewidth]{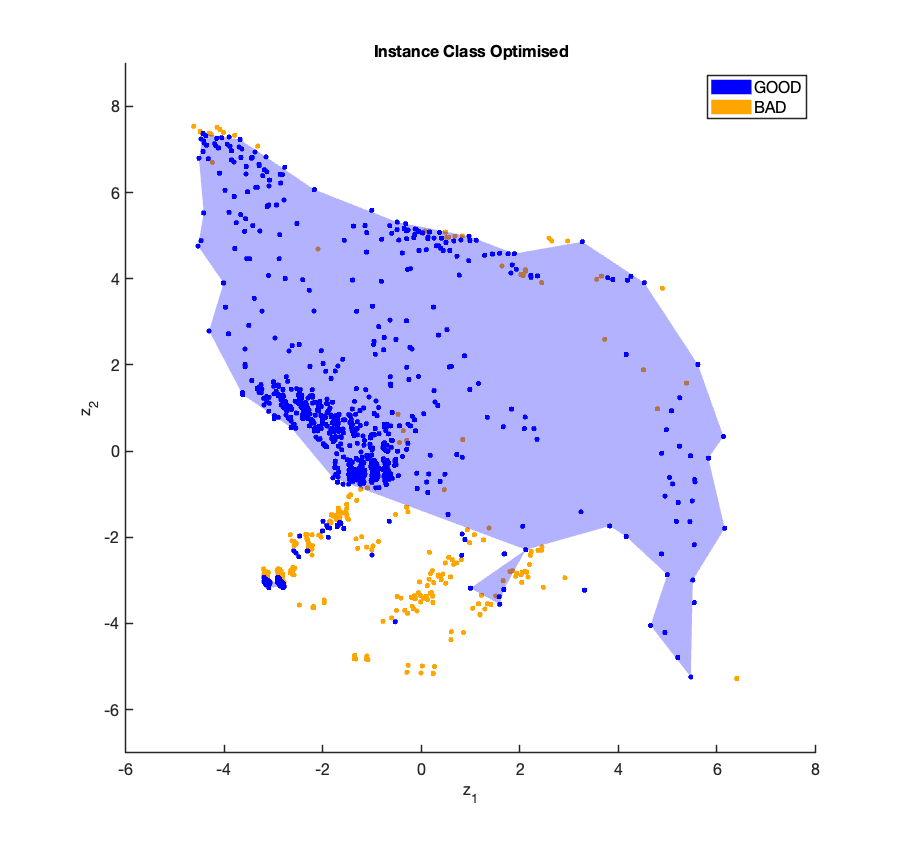}
\caption{Comparison of algorithm footprints by each algorithm: random initialisation (top-left), TQA (top-right), 3-regular graph optimised (bottom-left), and instance class optimised (bottom-right).}
\label{fig:isa-footprint-by-algo}
\end{figure}

\end{APPENDIX}


\bibliographystyle{informs2014} 
\bibliography{Bibliography} 

\end{document}